\documentclass[hyper]{JHEP3}
\title{Solitons, kinks and extended hadron model based on the generalized sine-Gordon theory}
\author{Harold Blas \\Departamentos de Matem\'atica e F\'{\i}sica - ICET\\
Universidade Federal de Mato Grosso\\
 Av. Fernando Correa, s/n, Coxip\'o \\
78060-900, Cuiab\'a - MT - Brazil\\ E-mail: \email{blas@ufmt.br}}
\author{Hector L. Carrion\\
Instituto de F\'{\i}sica, Universidade de S\~ao Paulo,
\\
Caixa Postal 66318, 05315-970, S\~ao Paulo, SP, Brazil.\\
E-mail: \email{hlc@fma.if.usp.br}}  \abstract{The solitons and kinks of the
generalized $sl(3, \IC)$ sine-Gordon (GSG) model are explicitly
obtained through the hybrid of the Hirota and dressing methods in
which the {\sl tau} functions play an important role. The various
properties are investigated, such as the potential vacuum
structure, the soliton and kink solutions, and the soliton masses
formulae. As a reduced submodel we obtain the double sine-Gordon
model. Moreover, we provide the algebraic
  construction of the
  $sl(3, \IC)$ affine Toda model coupled to matter (Dirac spinor) (ATM) and through a gauge fixing procedure
  we obtain the
classical version of the generalized $sl(3, \IC)$ sine-Gordon
model (cGSG) which completely decouples from the Dirac spinors. In
the spinor sector we are left with Dirac fields coupled to cGSG
fields. Based on the equivalence between the $U(1)$ vector and
topological currents it is shown the confinement of the spinors
inside the solitons and kinks of the cGSG model providing an
extended hadron model for ``quark" confinement.}

\usepackage[dvips]{epsfig}

\def\br{\begin{eqnarray}}
\def\er{\end{eqnarray}}
\def\be{\begin{equation}}
\def\ee{\end{equation}}

\def\({\left(}
\def\){\right)}

\relax

\def\a{\alpha}

\def\b{\beta}

\def\d{\delta}
\def\D{\Delta}

\def\g{\gamma}
\def\G{\Gamma}

\def\l{\lambda}

\def\n{\nu}

\def\pa{\partial}

\def\ra{\rightarrow}
\def\s{\sigma}

\def\th{\theta}

\def\tp0{\Theta_{+}^{(0)}}
\def\tm0{\Theta_{-}^{(0)}}

\def\vp{\varphi}

\def\f#1#2#3 {f^{#1#2}_{#3}}

\def\win1{{\sf w_{1+\infty}}}

\def\Win1{{\sf W_{1+\infty}}}

\def\rlx{\relax\leavevmode}
\def\inbar{\vrule height1.5ex width.4pt depth0pt}
\def\IZ{\rlx\hbox{\sf Z\kern-.4em Z}}
\def\IR{\rlx\hbox{\rm I\kern-.18em R}}
\def\IC{\rlx\hbox{\,$\inbar\kern-.3em{\rm C}$}}
\def\IN{\rlx\hbox{\rm I\kern-.18em N}}
\def\IO{\rlx\hbox{\,$\inbar\kern-.3em{\rm O}$}}
\def\IP{\rlx\hbox{\rm I\kern-.18em P}}
\def\IQ{\rlx\hbox{\,$\inbar\kern-.3em{\rm Q}$}}
\def\IF{\rlx\hbox{\rm I\kern-.18em F}}
\def\IG{\rlx\hbox{\,$\inbar\kern-.3em{\rm G}$}}
\def\IH{\rlx\hbox{\rm I\kern-.18em H}}
\def\II{\rlx\hbox{\rm I\kern-.18em I}}
\def\IK{\rlx\hbox{\rm I\kern-.18em K}}
\def\IL{\rlx\hbox{\rm I\kern-.18em L}}
\def\one{\hbox{{1}\kern-.25em\hbox{l}}}
\def\0#1{\relax\ifmmode\mathaccent"7017{#1}%
B        \else\accent23#1\relax\fi}

                \def\JHEP#1#2#3{{\sl JHEP} {\bf#1} (#2) #3}

                \def\JMP#1#2#3{{\sl J. Math. Phys.} {\bf #1} (#2) #3}

                \def\RMP#1#2#3{{\sl Rev. Mod. Phys.} {\bf #1} (#2) #3}

                \def\JPA#1#2#3{{\sl J. Physics} {\bf A#1} (#2) #3}

                \def\a{\alpha}
                \def\b{\beta}

                \def\d{\delta}
                \def\D{\Delta}
                
                \def\g{\gamma}
                \def\G{\Gamma}
                \def\vp{\varphi}

                \def\/{\frac}

                \def\l{\lambda}

                \def\n{\nu}

                \def\pa{\partial}

                \def\ra{\rightarrow}
                \def\rh{\rho}
                \def\vp{\varphi}
                
                \def\s{\sigma}
                
                \def\th{\theta}

                \def\({\Big(}
                \def\){\Big)}
                \def\[{\Big[}
                \def\]{\Big]}

                \def\rlx{\relax\leavevmode}
                \def\inbar{\vrule height1.5ex width.4pt depth0pt}
                \def\IZ{\rlx\hbox{\sf Z\kern-.4em Z}}
                \def\IR{\rlx\hbox{\rm I\kern-.18em R}}
                \def\IC{\rlx\hbox{\,$\inbar\kern-.3em{\rm C}$}}
                \def\IN{\rlx\hbox{\rm I\kern-.18em N}}
                \def\IO{\rlx\hbox{\,$\inbar\kern-.3em{\rm O}$}}
                \def\IP{\rlx\hbox{\rm I\kern-.18em P}}
                \def\IQ{\rlx\hbox{\,$\inbar\kern-.3em{\rm Q}$}}
                \def\IF{\rlx\hbox{\rm I\kern-.18em F}}
                \def\IG{\rlx\hbox{\,$\inbar\kern-.3em{\rm G}$}}
                \def\IH{\rlx\hbox{\rm I\kern-.18em H}}
                \def\II{\rlx\hbox{\rm I\kern-.18em I}}
                \def\IK{\rlx\hbox{\rm I\kern-.18em K}}
                \def\IL{\rlx\hbox{\rm I\kern-.18em L}}
                \def\one{\hbox{{1}\kern-.25em\hbox{l}}}
                \def\0#1{\relax\ifmmode\mathaccent"7017{#1}%
                B        \else\accent23#1\relax\fi}
                
\def\tone{\hat{\tau}_0}
\def\ttwo{\hat{\tau}_1}
\def\tthree{\hat{\tau}_2}
\def\tfour{{\tilde \tau}_R^{\a_1}}
\def\tfive{{\tilde \tau}_R^{\a_2}}
\def\tsix{\tau_R^{\a_3}}
\def\tseven{\tau_L^{\a_1}}
\def\teight{\tau_L^{\a_2}}
\def\tnine{{\tilde \tau}_L^{\a_3}}
\def\tten{{\tilde \tau}_{L,(2)}^{\a_1}}
\def\televen{{\tilde \tau}_{L,(0)}^{\a_1}}
\def\ttwelve{{\tilde \tau}_{L,(1)}^{\a_2}}
\def\tthirteen{{\tilde \tau}_{L,(0)}^{\a_2}}
\def\tfourteen{\tau_{L,(1)}^{\a_3}}
\def\tfifteen{\tau_{L,(2)}^{\a_3}}
\def\tsixteen{\tau_{R,(0)}^{\a_1}}
\def\tseventeen{\tau_{R,(2)}^{\a_1}}
\def\teighteen{\tau_{R,(0)}^{\a_2}}
\def\tnineteen{\tau_{R,(1)}^{\a_2}}
\def\ttwenty{{\tilde \tau}_{R,(1)}^{\a_3}}
\def\ttwentyone{{\tilde \tau}_{R,(2)}^{\a_3}}

\def\CSF#1#2#3{{\sl Chaos, Solitons and Fractals} {\bf C#1} (#2) #3}

                \def\JHEP#1#2#3{{\sl JHEP} {\bf#1} (#2) #3}

                \def\JMP#1#2#3{{\sl J. Math. Phys.} {\bf #1} (#2) #3}

                \def\RMP#1#2#3{{\sl Rev. Mod. Phys.} {\bf #1} (#2) #3}

                \def\JPA#1#2#3{{\sl J. Physics} {\bf A#1} (#2) #3}

\def\PD#1#2#3{{\sl Physica} {\bf D#1} (#2) #3}

\def\JPG#1#2#3{{\sl J. Phys.} {\bf G#1} (#2) #3}

\def\rme{\mathrm e}
\def\iu{\mathrm i\,}

                \def\a{\alpha}
                \def\b{\beta}

                \def\d{\delta}
                \def\D{\Delta}
                
                \def\g{\gamma}
                \def\G{\Gamma}
                \def\vp{\varphi}

                \def\/{\frac}

                \def\l{\lambda}

                \def\n{\nu}

                \def\pa{\partial}

                \def\ra{\rightarrow}
                \def\rh{\rho}
                \def\vp{\varphi}
                
                \def\s{\sigma}
                
                \def\th{\theta}

                \def\({\Big( }
                \def\){\Big)}
                \def\[{\Big[}
                \def\]{\Big]}
  \def\rlx{\relax\leavevmode}
                \def\inbar{\vrule height1.5ex width.4pt depth0pt}
                \def\IZ{\rlx\hbox{\sf Z\kern-.4em Z}}
                \def\IR{\rlx\hbox{\rm I\kern-.18em R}}
                \def\IC{\rlx\hbox{\,$\inbar\kern-.3em{\rm C}$}}
                \def\IN{\rlx\hbox{\rm I\kern-.18em N}}
                \def\IO{\rlx\hbox{\,$\inbar\kern-.3em{\rm O}$}}
                \def\IP{\rlx\hbox{\rm I\kern-.18em P}}
                \def\IQ{\rlx\hbox{\,$\inbar\kern-.3em{\rm Q}$}}
                \def\IF{\rlx\hbox{\rm I\kern-.18em F}}
                \def\IG{\rlx\hbox{\,$\inbar\kern-.3em{\rm G}$}}
                \def\IH{\rlx\hbox{\rm I\kern-.18em H}}
                \def\II{\rlx\hbox{\rm I\kern-.18em I}}
                \def\IK{\rlx\hbox{\rm I\kern-.18em K}}
                \def\IL{\rlx\hbox{\rm I\kern-.18em L}}
\preprint{}

\begin{document}

\section{Introduction}

The sine-Gordon model (SG) has been studied over the decades due
to its many properties and mathematical structures such as
integrability and soliton solutions. It can be used as a toy model
for non-perturbative quantum field theory phenomena. In this
context, some extensions and modifications of the SG model deserve
attention. An extension taking multi-frequency
 terms as the potential has been investigated in connection to various physical applications
  \cite{delfino, bajnok, sodano, mussardo}.

On the other hand, an extension defined for multi-fields is the
so-called generalized sine-Gordon model (GSG) which has been found
in the
 study of the strong/weak coupling sectors of the so-called $sl(N,
\IC)$ affine Toda coupled to matter fields (ATM) theory \cite{jmp,
jhep}. In connection to these developments, the bosonization
process of the multi-flavor massive Thirring model (GMT)  provides
the quantum version of the (GSG) model \cite{epjc}. The GSG model
 provides a framework to obtain (multi-)soliton solutions for unequal
mass parameters of the fermions in the GMT sector and study the
spectrum and their interactions. The extension of this picture to
the NC space-time has been addressed (see \cite{jhep12} and
references therein).

Coupled systems of scalar fields have been investigated by many
authors \cite{Rajaraman, riazi, pogosian, Izquierdo, Bazeia1,
Bazeia2}. One of the motivations was the study of topological
defects in relativistic field theories; since realistic theories
involve more than one scalar field, the multi-field sine-Gordon
 theories with kink-type exact solutions deserve some attention.
The interest in the study of the classical limit of string theory
on determined backgrounds has recently been greatly stimulated in
connection to integrability. It has been established that the
classical string on $R \times S^2$ is essentially equivalent to
the sine-Gordon integrable system \cite{KP}. More recently, on $R
\times S^3$ background utilizing the Pohlmeyer's reduction it has
been obtained a family of classical string solutions called dyonic
giant magnons which were associated with solitons of complex
sine-Gordon equations \cite{okamura}. String theory on $R\times
S^{N-1}$ is classically equivalent to the so-called $SO(N)$
symmetric space sine-Gordon model (SSG)
 \cite{mikhailov2}.

In this paper we study the spectrum of solitons and kinks  of the
GSG model proposed in \cite{jmp, jhep, epjc} and consider the
closely related ATM model from which one gets the classical GSG
model (cGSG) through a gauge fixing procedure. Some reductions of
the GSG model to one-field theory lead to the usual SG model and
to the so-called multi-frequency sine-Gordon models. In
particular, the double (two-frequency) sine-Gordon model (DSG)
appears in a reduction of the $sl(3, \IC)$ GSG model. The DSG
theory is a non–integrable quantum field theory with many physical
applications \cite{sodano, mussardo}.

Once a convenient gauge fixing is performed by setting to constant
some spinor bilinears in the ATM model we are left with two
sectors: the cGSG model which completely decouples from the
spinors and a system of Dirac spinors coupled to the cGSG fields.
Following the references \cite{chang, Uchiyama} in which a
$1+1$-dimensional bag model for quark confinement is considered,
we follow their ideas and generalize for multi-flavor Dirac
spinors coupled to cGSG solitons and kinks. The first reference
considers a model similar to the $sl(2)$ ATM theory, and in the
second one the DSG kink is proposed as an extended hadron model.

In the next section we define the $sl(3, \IC)$ GSG model and study
its properties such as the vacuum structure and the soliton, kink
 and bounce type solutions. In section \ref{atm} we consider the
 $sl(3,\IC)$ affine Toda model coupled to matter and obtain the cGSG
 model through a gauge fixing procedure. It is discussed the physical soliton
spectrum of the gauge fixed model. In section \ref{topological}
the topological charges are introduced, as well as the idea of
baryons as solitons (or kinks), and the quark confinement
mechanism is discussed. The discussion section outlines possible
directions for future research, in particular, the GSG application
to QCD$_{2}$. In appendix \ref{atmapp} we provide the zero
curvature formulation of the $sl(3,\IC)$ ATM model.

\section{The model}
\label{model}

The generalized sine-Gordon model (GSG) related to $sl(N, \IC)$ is
defined  by \cite{jmp, jhep, epjc} \br \label{GSG} S= \int d^2x
\sum_{i=1}^{N_{f}}\[ \frac{1}{2} (\pa_{\mu} \Phi_{i})^2 + \mu_{i}
\( \mbox{cos} \b_{i} \Phi_{i}-1\)\]. \er The $\Phi_{i}$ fields in
(\ref{GSG})
 satisfy the constraints \br \label{constr0}  \Phi_{p}= \sum_{i=1}^{N-1}
\s_{p\,i} \Phi_{i},\,\,\,\,\,p=N,N+1,...,
N_{f},\,\,\,\,N_{f}=\frac{N(N-1)}{2},\er where
 $\s_{p\,i}$
are some constant parameters and $N_{f}$ is the number of positive
roots of the Lie algebra $sl(N, \IC)$. In the context of the Lie
algebraic construction of the GSG system these constraints arise
from the relationship between the positive and simple roots of
$sl(N, \IC)$. Thus, in (\ref{GSG}) we have $(N-1)$ independent
fields related to the number of simple roots of the $sl(N, \IC)$
Lie algebra.

We consider the $sl(3, \IC)$ algebra, since in this case one has
two simple roots there are two independent real fields, $\vp_{1,
\,2}$, associated to them such that \br \label{fields} \Phi_{1}=
2\vp_{1}-\vp_{2};\,\,\,\Phi_{2}= 2\vp_{2}-\vp_{1};\,\,\,\Phi_{3}=
r\, \vp_{1}+ s\, \vp_{2},\,\,\,\,s,r \in \IR \er which must
satisfy the constraint
\begin{eqnarray} \label{constr} \b_{3}
\Phi_{3}= \d_{1} \b_{1} \Phi_{1}+\d_{2} \b_{2}
\Phi_{2},\,\,\,\,\b_{i}\equiv \b_{0}\nu_{i},
\end{eqnarray}
where $\b_{0},\,\nu_{i},\,\d_{1}, \d_{2}$ are some real numbers.
The $\Phi$ fields dependence on the $\vp$ ones
 will be explained in the context of the Lie algebraic
construction of the classical version of the model in sections
\ref{atm} and \ref{topological}. In view of the definitions above,
the $sl(3, \IC)$ GSG model can be regarded as three usual
sine-Gordon models coupled through the linear constraint
(\ref{constr}).

Taking into account (\ref{fields})-(\ref{constr}) and the fact
that the fields $\vp_{1}$ and $\vp_{2}$ are independent we may get
the relationships \br\label{nus} \nu_{2} \d_{2} = \rho_{0} \nu_{1}
\d_{1} \,\,\,\,\,\nu_{3} =\frac{1}{r+s}(\nu_{1} \d_{1}+\nu_{2}
\d_{2} );\,\,\,\, \rho_{0} \equiv \frac{2s+r}{2r+s} \er

The $sl(3, \IC)$ model has a potential density \br V[\vp_{i}] =
\sum_{i=1}^{3} \mu_{i}\(1- \mbox{cos} \b_{i}\Phi_{i}\)
\label{potential}\er

The GSG model has been found in the process of bosonization of the
generalized massive Thirring model (GMT) \cite{epjc}. The GMT
model is a multiflavor extension of the usual massive Thirring
model incorporating massive fermions with current-current
interactions between them. In the $sl(3, \IC)$ construction of
\cite{epjc} the parameters $\d_{i}$ depend on the couplings
$\b_{i}$ and  they satisfy certain relationship. This is obtained
by assuming $\mu_{i}
>0$ and the zero of the potential given for $\Phi_{i}=
\frac{2\pi}{\b_{i}} n_{i}$, which substituted into (\ref{constr}) provides \br
\label{deltas} n_{1} \d_{1}+ n_{2} \d_{2}= n_{3},\,\,\,\,n_{i} \in \IZ \er

The last relation combined with (\ref{nus}) gives \br (2r+s)
\frac{n_{1}}{\nu_{1}}+ (2s+r) \, \frac{n_{2}}{\nu_{2}} = 3
\,\frac{n_{3}}{\nu_{3}}\label{nns}.\er

The periodicity of the potential implies an infinitely degenerate
ground state and then the theory supports topologically charged
excitations. A typical potential is plotted in Fig. 1. The vacuum
configuration is related to the fundamental weights (see sections
\ref{atm}, \ref{topological} and the Appendix). For the moment,
consider the fields $\Phi_{1}$ and $\Phi_{2}$ and the vacuum
lattice defined by \br (\Phi_{1}\,, \, \Phi_{2})
=\frac{2\pi}{\b_{0}} (\frac{n_{1}}{\nu_{1}}
\,,\,\frac{n_{2}}{\nu_{2}}),\,\,\,\,\,n_{a} \in \IZ.
\label{lattice1}\er

\FIGURE{\centering
\hspace{2.0cm}\scalebox{0.8}{\includegraphics{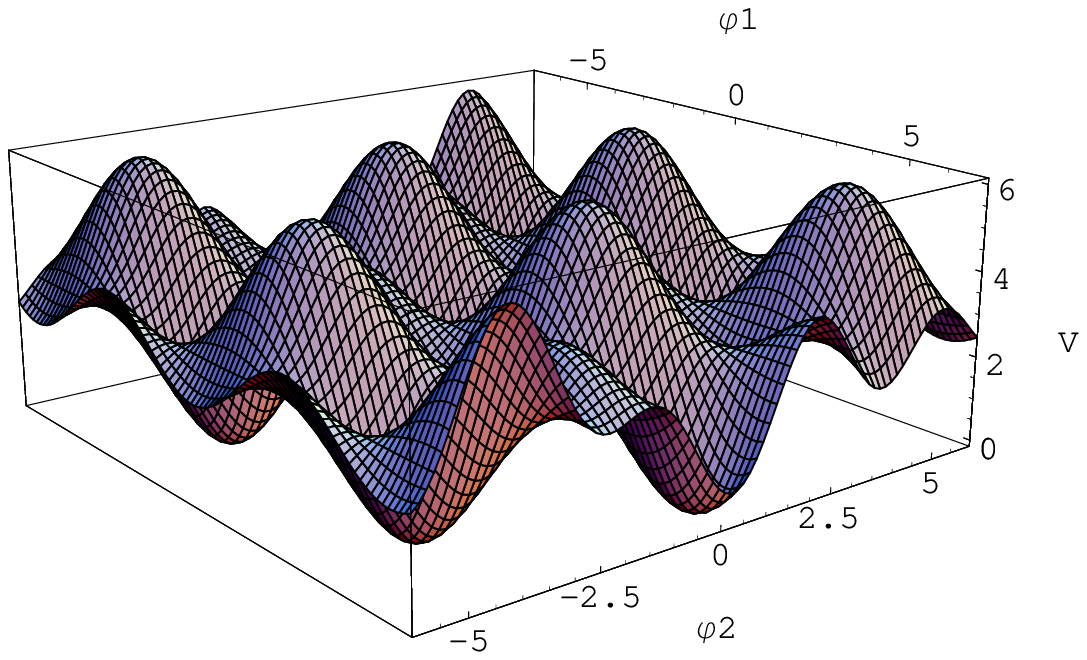}}
\parbox{5in}{\caption{GSG potential $V$ for the parameter
values $\nu_{1}=1/2,\,\,
\delta_{1}=2,\,\,\delta_{2}=1,\,\,\nu_{2}=1,\,\,
r=s=1,\,\,\b_{0}=1,\,\,\mu_{1}=\mu_{2}=1$.}}}

It is convenient to write the equations of motion in terms of the
independent fields $\vp_{1}$ and $\vp_{2}$  \br \pa^2 \vp_{1} &=&
- \mu_{1} \b_{1} \D_{11} \mbox{sin} [\b_{1} ( 2\vp_{1}-\vp_{2})]-
\mu_{2} \b_{2} \D_{12} \mbox{sin}[\b_{2} (2\vp_{2}-\vp_{1})]+
\nonumber\\&& \mu_{3} \b_{3} \D_{13} \mbox{sin} [\b_{3} ( r
\vp_{1} + s \vp_{2})]
\label{eq1}\\
\pa^2 \vp_{2} & = & - \mu_{1} \b_{1} \D_{21} \mbox{sin} [\b_{1} (
2\vp_{1}-\vp_{2})]-\mu_{2} \b_{2} \D_{22} \mbox{sin}[\b_{2}
(2\vp_{2}-\vp_{1})]+ \nonumber\\  && \mu_{3} \b_{3} \D_{23}
\mbox{sin} [\b_{3} ( r \vp_{1}+ s \vp_{2})]\label{eq2}, \er where
\br \nonumber A&=&\b_{0}^{2}\nu_{1}^2(4+\d^2+ \d_{1}^2\rh_{1}^2
r^2),\,\,\,\,B=\b_{0}^{2}\nu_{1}^2 (1+4 \d^2 + \d_{1}^2\rh_{1}^2
s^2),\,\,\,\,C=\b_{0}^{2}\nu_{1}^2(2+2 \d^2+ \d_{1}^2\rh_{1}^2 r\,
s),\\\nonumber
\D_{11}&=&(C-2B)/\D,\,\,\,\,\D_{12}=(B-2C)/\D,\,\,\,\,\D_{13}=(r\,
B+s\, C)/\D,\\\nonumber
\D_{21}&=&(A-2C)/\D,\,\,\,\,\D_{22}=(C-2A)/\D,\,\,\,\,\D_{23}=(r\,C+s\,A)/\D\\
 \D&=&C^2-AB,\,\,\,\,\d=\frac{\d_{1}}{\d_{2}} \rh_{0},\,\,\,\,\rh_{1}=\frac{3}{2r+s}\nonumber \er

Notice that the eqs. of motion (\ref{eq1})-(\ref{eq2}) exhibit the
symmetries \br\label{symm} \vp_{1} &\leftrightarrow &
\vp_{2},\,\,\,\,\mu_{1} \leftrightarrow \mu_{2}, \,\,\,\nu_{1}
\leftrightarrow \nu_{2},\,\,\,\d_{1} \leftrightarrow \d_{2},\,\,\,
r \leftrightarrow s; \\
&\mbox{and}&\,\,\,\vp_{a} \leftrightarrow - \vp_{a},\,\,\,a=1,2
\label{changesign}\er

Some type of coupled sine-Gordon models have been considered in
connection to various interesting physical problems
\cite{kivshar}. For example a system of two coupled SG models has
been proposed in order to describe the dynamics of soliton
excitations in deoxyribonucleic acid (DNA) double helices
\cite{zhang}. In general these type of equations have been solved
by perturbation methods around decoupled sine-Gordon exact
solitons.

The system of equations (\ref{eq1})-(\ref{eq2}) for  certain
choice of the parameters $r$  and $s$ will be derived in section
\ref{atm} in the context of the $sl(3)$ ATM type models, in which
the fields $\vp_{1}$ and $\vp_{2}$ couple to some Dirac spinors in
such a way that the model exhibits a local gauge invariance. The
ATM relevant equations of motion have been solved using a hybrid
of the Hirota and Dressing methods \cite{bueno}. However, in this
reference the physical spectrum of solitons and kinks of the
theory, related to a convenient gauge fixing of the model, have
not been discussed, even though the topological and Noether
currents equivalence has been verified. The appearance of the
so-called tau functions, in order to find soliton solutions in
integrable models, is quite a general result in the both Dressing
and Hirota approaches. In this section,  we will find soliton and
kink type solutions of the GSG model (\ref{eq1})-(\ref{eq2}) and
closely follow the spirit of the above hybrid method approach to
find soliton solutions.

The general tau function for an $n-$soliton solution of the {\sl
gauge unfixed} ATM model has the form \cite{bueno, ferreira} \br
\label{tau} \tau &=& \sum_{p_{1}.....p_{n}=0}^{2}
c_{p_{1}.....p_{n}} \mbox{exp}[p_{1} \G_{i_{1}}(z_{1})+...+p_{n}
\G_{i_{n}}(z_{n})],\,\,\,\,  z_{i}= \g_{i}(x-v_{i}\,
t),\\
\nonumber && c_{p_{1}.....p_{n}}\in \IC  \er

Since the GSG model describes the strong coupling sector (soliton
 spectrum) of the ATM model \cite{jmp, jhep} then one can guess the following
Ansatz for the tau functions of the GSG model \br \label{taus}
e^{-i\b_{0}
\frac{\vp_{1}}{2}}=\frac{\tau_{1}}{\tau_{0}},\,\,\,\,\,e^{-i\b_{0}
\frac{\vp_{2}}{2}}=\frac{\tau_{2}}{\tau_{0}}, \er where the tau
functions $\tau_{i}\,(i=0,1,2)$ are assumed to be of the form
(\ref{tau}). The relationships (\ref{taus}) mimic the ones
appearing in section \ref{atm} below for the relations between the
ATM scalar fields and the relevant three tau functions (see Eq.
(\ref{phitau})). We have two independent fields $\vp_{1,\,2}$ in
the model (\ref{eq1})-(\ref{eq2}). However, in (\ref{taus}) we
have introduced three tau-functions. The one equation missing
arises from an algebraic relation between the tau functions
$\tau_{1,\,2}$. In fact, from the relationships (\ref{fields}),
(\ref{constr}) and (\ref{nus}), together with (\ref{taus}) one
deduces that the tau functions $\tau_{1,\,2}$ must be of the form
\br \tau_{a} &=& |\tau|
\,e^{i\zeta_{a}};\,\,\,\,(a=1,2),\,\,\,\,\, \zeta_{1}-\zeta_{2}=
\frac{2\pi
n}{k},\label{zetas}\\
&& k=2[(2r+s)\d_{2}\nu_{2}-(2s+r)\d_{1}\nu_{1}],\,\,\,n\in \IZ .
\nonumber\er

Assuming that the fields $\vp_{a}$ are real one has that (since
$-i \vp_{a}=\mbox{log} |\frac{\tau_{a}}{\tau_{0}}|+ i \mbox{arg}
 \frac{\tau_{a}}{\tau_{0}}$) \br \label{modulus} |\tau_{0}|=|\tau_{a}|=|\tau|,\,\,\,(a=1,2)\er
where $|\tau|$ has been defined in (\ref{zetas}).

 Therefore, we are
parameterizing two independent real fields $\vp_{1,2}$ with three
tau functions satisfying the constraints (\ref{zetas}) and
(\ref{modulus}). From (\ref{taus}) one can write \br
\vp_{1,\,2}&=&\frac{4}{\b_{0}} \mbox{arctan}\left[\frac{e_{1}
\([Re(\tau_{1,\,2})]^2+[Im(\tau_{1,\,2})]^2\)-\(Re(\tau_{1,\,2})
Re(\tau_{0})+Im(\tau_{1,\,2}) Im(\tau_{0})\)}{\[Im(\tau_{1,\,2})
Re(\tau_{0})-
 Re(\tau_{1,\,2}) Im(\tau_{0})\]}\right],\nonumber \\
 \label{arct}\\\nonumber
e_{1}&=&\pm 1 \label{e1},\,\,\,\,\mbox{or}\\
\vp_{a} &=& \zeta_{0}-\zeta_{a} + n_{a} \pi,\,\,\,n_{a}\in
\IZ,\,\,\,\,a=1,2;\er where the phases $\zeta_{1, 2}$ were defined
in (\ref{zetas}), and $\zeta_{0}$ corresponds to the definition
$\tau_{0}= |\tau_{0}| e^{i \zeta_{0}}$.

According to the symmetry (\ref{changesign}) of the system of Eqs.
(\ref{eq1})-(\ref{eq2}) there will be another solution written as
\br \label{taus1} e^{-i\b_{0}
\frac{\vp_{1}^{'}}{2}}=\frac{\tau_{0}}{\tau_{1}},\,\,\,\,\,e^{-i\b_{0}
\frac{\vp_{2}^{'}}{2}}=\frac{\tau_{0}}{\tau_{2}}, \er for the same
tau functions $\tau_{i}\,(i=0,1,2)$.

We will see that the Ansatz (\ref{taus}) provides soliton and kink
type solutions of the model (\ref{eq1})-(\ref{eq2}), in this way
justifying {\sl a posteriori} the assumption made for the number
and form of the tau functions.

According to the Hirota method one substitutes the relations
between the fields and tau functions (\ref{taus}) into the
equations of motion (\ref{eq1})-(\ref{eq2}), so in terms of the
tau functions these equations become \br \frac{2i}{\b^2_{0}}\[
\frac{\pa^2 \tau_{1}}{\tau_{1}}-\frac{(\pa
\tau_{1})^2}{\tau_{1}^2}-\frac{\pa^2 \tau_{0}}{\tau_{0}}
+\frac{(\pa \tau_{0})^2}{\tau_{0}^2}\]+
 \frac{\b_{1}\mu_{1} \D_{11}}{2i} \[ \frac{(\tau_{2} \tau_{0})^{4\nu_{1}}-\tau_{1}^{8\nu_{1}}}{(\tau_{2} \tau_{0})^{2\nu_{1}} \tau_{1}^{4\nu_{1}}}\]&+&\nonumber \\
 \frac{\b_{2}\mu_{2} \D_{12}}{2i} \[ \frac{(\tau_{1} \tau_{0})^{4\nu_{2}}-\tau_{2}^{8\nu_{2}}}{(\tau_{1} \tau_{0})^{2\nu_{2}} \tau_{2}^{4\nu_{2}}}\]-
 \frac{\b_{3}\mu_{3} \D_{13}}{2i} \[ \frac{(\tau_{0})^{4\nu_{3}(r+s)}-\tau_{1}^{4 r\nu_{3}}\tau_{2}^{4 s\nu_{3}}}{(\tau_{2})^{2s\nu_{3}} (\tau_{1})^{2r\nu_{3}}
 \tau_{0}^{2\nu_{3}(r+s)}}\]&=&0,
 \label{eq1tau}\\
\frac{2i}{\b^2_{0}}\[ \frac{\pa^2 \tau_{2}}{\tau_{2}}-\frac{(\pa \tau_{2})^2}{\tau_{2}^2}-\frac{\pa^2 \tau_{0}}{\tau_{0}}
+\frac{(\pa \tau_{0})^2}{\tau_{0}^2}\]+
 \frac{\b_{1}\mu_{1} \D_{21}}{2i} \[ \frac{(\tau_{2} \tau_{0})^{4\nu_{1}}-\tau_{1}^{8\nu_{1}}}{(\tau_{2} \tau_{0})^{2\nu_{1}} \tau_{1}^{4\nu_{1}}}\]&+&\nonumber \\
 \frac{\b_{2}\mu_{2} \D_{22}}{2i} \[ \frac{(\tau_{1} \tau_{0})^{4\nu_{2}}-\tau_{2}^{8\nu_{2}}}{(\tau_{1} \tau_{0})^{2\nu_{2}} \tau_{2}^{4\nu_{2}}}\]-
 \frac{\b_{3}\mu_{3} \D_{23}}{2i} \[ \frac{(\tau_{0})^{4\nu_{3}(r+s)}-\tau_{1}^{4 r\nu_{3}}\tau_{2}^{4 s\nu_{3}}}{(\tau_{2})^{2s\nu_{3}} (\tau_{1})^{2r\nu_{3}}
 \tau_{0}^{2\nu_{3}(r+s)}}\]&=&0.
 \label{eq2tau}
\er

These equations will be used to implement a computer program for
algebraic manipulations like MAPLE in order to verify the soliton
type solutions provided the relevant tau functions are supplied.
We will see that the 1-soliton and 1-kink type solutions are
related to half-integer or integer values of the parameters
$\nu_{i}$ and the values $r, s \,=\, 0, 1$. In the next
subsections we will write the 1-soliton(antisoliton),
1-kink(antikink) and bounce type solutions.

\subsection{One soliton/antisoliton pair associated to $\vp_{1}$}
\label{s11}

Consider the tau functions \br \tau_{0}&=& 1+ i \,d\,
\mbox{exp}[\gamma (x-v t)];\,\,\,\,\tau_{1}=1- i\, d\,
\mbox{exp}[\gamma (x-v t)];\,\,\,\,\\ \tau_{2}&=& 1+ i\, d\,
\mbox{exp}[\gamma (x-v t)].\er

This choice satisfies the system of equations
(\ref{eq1tau})-(\ref{eq2tau}) for the set of parameters \br
\label{para1} \nu_{1}=1/2,\,\,
\delta_{1}=2,\,\,\delta_{2}=1,\,\,\nu_{2}=1,\,\,
\nu_{3}=1,\,\,r=1.\er provided that \br
13\mu_{3}=5\mu_{2}-4\mu_{1},\,\,\,\,
\gamma^2_{1}=\frac{1}{13}(6\mu_{2}+3\mu_{1}).\er

Now, taking  $e_{1}=1$ in Eq. (\ref{e1}) and the relation (\ref{arct}) one has
\br \label{sol1}
\vp_{1}= -\frac{4}{\b_{0}}\mbox{arctan}\{d\,\, \mbox{exp}[\gamma_{1}
(x-v t)]\},\,\,\,\,\vp_{2}=0. \er

This solution is precisely the sine-Gordon 1-antisoliton
associated to the field $\vp_{1}$ with mass $M_{1}=\frac{8
\g_{1}}{\b^2_{0}}$. This solution corresponds to the relations
(\ref{taus1}). We plot an antisoliton of this type in Fig. 3.

\subsection{One soliton/antisoliton pair associated to $\vp_{2}$}
\label{s22}

Next, let us consider the tau functions \br \tau_{0}&=&1+i\,
d\,\mbox{exp}[\gamma(x-vt)],\,\,\,\,\tau_{1}=
1+i\,d\,\mbox{exp}[\gamma
(x-vt)],\\
\tau_{2}&=&1-i\,d\,\mbox{exp}[\gamma (x-vt)]\er

This set of tau functions solves the system
(\ref{eq1tau})-(\ref{eq2tau}) for the choice of parameters \br
\label{para2} \nu_{1}=1,\,\,
\delta_{1}=1,\,\,\delta_{2}=2,\,\,\nu_{2}=1/2,\,\,
\nu_{3}=1,\,\,s=1\er provided that \br
13\mu_{3}=5\mu_{1}-4\mu_{2},
\,\,\,\,\gamma^2_{2}=\frac{1}{13}(6\mu_{1}+3\mu_{2}) \er

Now, choose $e_{1}=1$ in (\ref{e1}) and through (\ref{arct}) one can get
\br \label{sol2} \vp_{2}= -\frac{4}{\b_{0}}
\mbox{arctan}\{d\,\mbox{exp}[\gamma_{2}(x-vt)]\},\,\,\,\,\vp_{1}=0\er

Similarly, this is the sine-Gordon 1-antisoliton associated to the
field $\vp_{2}$ with mass $M_{2}=\frac{8 \g_{2}}{\b^2_{0}}$ and
its profile is of the type shown in Fig 3. Likewise, this solution
corresponds to the relations (\ref{taus1}).

\subsection{Two 1-soliton/1-antisoliton pairs associated to $\hat{\vp}_{A} \equiv \vp_{1}=\vp_{2}\,\,(A=1,2)$}
\label{s33}

Now, let us consider the tau functions \br \tau_{0}& =& 1+
i\,d\,\,\mbox{exp}[\gamma(x-v t)],\,\,\,\, \tau_{1} = 1 -
i\,d\,\,\mbox{exp}[\gamma(x-v t)],\\
\tau_{2} &=& 1 - i\,d\,\,\mbox{exp}[\gamma(x-v t)].\er

This choice satisfies (\ref{eq1tau})-(\ref{eq2tau}) for
 \br \label{para3} \nu_{1}=1,\, \,\delta_{1}=1/2,\,\,\nu_{2}=1,\,\,
 \delta_{2}=1/2,\,\, \nu_{3}=1/2,\,\,r=s=1,\er  provided that
\br
d^2=1, \,\,\,\,38\gamma^2_{3} = 25\mu_{1}+13\mu_{2}+19\mu_{3}
\er

Now, taking $e_{1}=1$ in (\ref{arct})  one has \br
\vp_{1}&=&\vp_{2}\equiv \hat{\vp}_{1},\,\,\,\,  \label{sol3a}
\\\hat{\vp}_{1} &=& -\frac{4}{\b_{0}} \mbox{arctan}\{d\,\,
\,\mbox{exp}[\gamma_{3}(x-vt)]\}. \label{sol3b}\er

This is a sine-Gordon 1-antisoliton associated to both fields
$\vp_{1,\,2}$ in the particular case when they are equal to each
other, and it corresponds to the relations (\ref{taus1}).  It
possesses a mass $M_{3}=\frac{8\g_{3}}{\b_{0}^2}$.

In view of the symmetry (\ref{symm}) we are able to write \br
d^2=1, \,\,\,\,38\gamma^2_{4} = 25\mu_{2}+13\mu_{1}+19\mu_{3}, \er
and then on has another soliton of this type \br
\vp_{1}&=&\vp_{2}\equiv \hat{\vp}_{2},\, \label{sol3c}
\\\hat{\vp}_{2}&=& -\frac{4}{\b_{0}} \mbox{arctan}\{d\,\, \,\mbox{exp}[\gamma_{4}(x-vt)]\}.
\label{sol3d}\er

It possesses a mass $M_{4}=\frac{8\g_{4}}{\b_{0}^2}$ and
corresponds to the relations (\ref{taus1}). This 1-antisoliton is
of the type shown in Fig. 3.

The GSG system (\ref{eq1})-(\ref{eq2}) reduces to the usual SG
equation for each choice of the parameters (\ref{para1}),
(\ref{para2}) and (\ref{para3}), respectively. Then, the
$n-$soliton solutions in each case can be constructed as in the
ordinary sine-Gordon model by taking appropriate tau functions in
(\ref{tau})-(\ref{taus}).

In view of the symmetry (\ref{changesign}) one can be able to
construct the {\sl solitons} corresponding to the antisoliton
solutions (\ref{sol1}), (\ref{sol2}), (\ref{sol3b}) and
(\ref{sol3d}) simply by changing their signs $\vp_{a} \rightarrow
-\vp_{a}$.

A modified model with rich soliton dynamics is the so-called
stepwise sine-Gordon model in which the system
  parameter depends on the sign of the SG field \cite{riazi1}. It
  would be interesting to consider the above GSG model along the
  lines of this reference.

\subsection{Mass splitting of solitons}
\label{splitt}

It is interesting to write some relations among  the various
soliton masses \br M_{3}^2= \frac{1}{76} (109 M_{2}^2 + 5
M_{1}^2);\,\,\,\, M_{4}^2= \frac{1}{76} (109 M_{1}^2 + 5 M_{2}^2);
\er

If $\mu_{1}=\mu_{2}$  then we have the degeneracy $M_{1}=M_{2}$, and $M_{3}=M_{4}=
\sqrt{3/2} M_{1}$. Notice that if $M_{1}\neq M_{2}$ then $M_{3} < M_{1}+ M_{2}$ and $M_{4} < M_{1}+ M_{2}$,
and the third and fourth solitons are stable in the sense that energy is required to dissociate them.

\subsection{Kinks of the reduced two-frequency sine-Gordon model}
\label{dsg:sec}

In the system (\ref{eq1})-(\ref{eq2}) we perform the following
reduction $\vp \equiv \vp_{1}=\vp_{2}$ such that \br
\Phi_{1}=\Phi_{2},\,\,\,\Phi_{3}= q \, \Phi_{1}, \label{reduc}\er
with $q$ being a real number. Therefore, using the constraint
(\ref{constr}) one can deduce the relationships \br \d_{1} =
\frac{q}{2},\,\,\,\d =1 \label{re1}. \er

Moreover, for consistency of the system of equations
(\ref{eq1})-(\ref{eq2}) we have to impose the relationships \br
\nu_{1}\mu_{1}\D_{11}+ \nu_{2}\mu_{2}\D_{12} &=&
\nu_{1}\mu_{1}\D_{21}+\nu_{2}\mu_{2}\D_{22},\\
\D_{13}&=&\D_{23}.\er

These relations imply \br \label{re2} \d^2 = 1, \,\,\,\mu_{1}=\d
\,\,\mu_{2} \er.

Taking into account the  relations (\ref{re1}) and (\ref{re2})
together with (\ref{nus}) we get \br \mu_{1}=\mu_{2},\,\,\,\,\d =1
,\,\,\,\nu_{1}=\nu_{2},\,\,\,\nu_{3}=\frac{q}{2}
\nu_{1},\,\,r=s=1. \label{param1}\er

Thus the system of Eqs.(\ref{eq1})-(\ref{eq2}) reduce to \br
\label{dsg}\pa^2 \Phi &=& -
\frac{\mu_{1}}{\nu_{1}}\,\mbox{sin}(\nu_{1} \Phi)-\frac{\mu_{3}
\d_{1}}{\nu_{1}} \mbox{sin} (q\, \nu_{1} \Phi),\,\,\,\,\, \Phi
\equiv \b_{0} \vp. \er

This is the so-called {\sl two-frequency (double) sine-Gordon}
model (DSG) and it has been the subject of much interest in the
last decades, from the mathematical and physical points of view.
It encounters many interesting physical applications, such as to
the study of massive Schwinger model (two-dimensional quantum
electrodynam- ics) and a generalized Ashkin-Teller model (a
quantum spin system).  A further potentially interesting
application of the two-(and multi-)frequency sine-Gordon model is
for ultra-short optical pulses propagating in resonant degenerate
medium (see e.g. \cite{sodano, mussardo, Uchiyama, kivshar}).

If the parameter $q$ satisfies \br \label{frac1} q =  \frac{n}{m}
\, \in \,  \IQ \er with $m, \,n$ being two relative prime positive
integers, then the potential
$\frac{\mu_{1}}{\nu_{1}^2}(1-\mbox{cos} (\nu_{1} \Phi))
+\frac{\mu_{3}}{2 \nu_{1}^2}(1-\mbox{cos} (q \nu_{1} \Phi))$
associated to the model (\ref{dsg}) is periodic with period \br
\frac{2\pi}{\nu_{1} } m = \frac{2\pi}{q\, \nu_{1} } n. \er

As mentioned above the theory (\ref{dsg}) possesses topological excitations.
The fundamental topological excitations degenerates in the $\mu_{1}=0$
limit to an $n-$soliton state of the relevant sine-Gordon model and similarly
in the limit $\mu_{3}=0$ it will be  an $m$-soliton state. For general values
of the parameters $\mu_{1},\,\mu_{3},\, \d_{1},\,\nu_{1}$ the solitons are in some sense
 ``confined'' inside the topological excitations which become in this form some composite
  objects. On the other hand, if $q \notin \IQ $ then the potential is not periodic,
  so, there are no topologically charged excitations and the solitons are completely
confined \cite{delfino, bajnok}.

The model (\ref{dsg}) in the limit $\mu_{1}=0$  reduces to
\br
\label{eqmu10}
\pa^2 \vp = - \frac{\mu_{3} q}{2 \nu_{1} \b_{0}}\, \mbox{sin} (q \nu_{1} \b_{0} \vp).
\er
For later discussion  we record here the mass of the soliton associated to this equation, \br M_{\mu_{3}} = \frac{8}{(q \nu_{1}\b_{0})^2} \sqrt{q^2 \mu_{3}/2}\label{massmu3}.\er

Correspondingly in the limit  $\mu_{3}=0$ one has \br
\label{eqmu30} \pa^2 \vp = - \frac{mu_{1}}{\nu_{1} \b_{0}}\,
\mbox{sin} (\nu_{1} \b_{0} \vp) \er with associated soliton mass
\br M_{\mu_{1}} = \frac{8}{(\nu_{1}\b_{0})^2} \sqrt{
\mu_{1}}\label{massmu1}\er

Notice that other possibilities to perform  the reduction of type
(\ref{reduc}) encounter some inconsistencies, e.g. the attempt to
implement the reduction $\Phi_{1}=\Phi_{3},\,\,\Phi_{2}= q' \,
\Phi_{1}$ implies $\d_{1,\,2}^2 < 0$ which is a contradiction
since $\d_{1,\,2}$ are real numbers by definition. The same
inconsistency occurs when one tries to reduce the $sl(3, \IC)$ GSG
model to a three-frequency SG model. We expect that the three and
higher frequency models \cite{toth} will be related to $sl(N,
\IC),\,N\ge 4,$ GSG models.

In the following we will provide some kink solutions for
particular set of parameters. Consider \br
\label{paras}\nu_{1}=1/2,\,\,
\d_{1}=\d_{2}=1,\,\,\nu_{2}=1/2,\,\,\,\nu_{3}=1/2 \,\,\,
\mbox{and} \,\,\, q=2,\,n=2,\, m=1\er which satisfy (\ref{param1})
and (\ref{frac1}), respectively. This set of parameters provide
the so-called {\sl double sine-Gordon model} (DSG). Its potential
$-[4 \mu_{1}(\mbox{cos} \frac{\Phi}{2}-1 )+ 2\mu_{3}(\mbox{cos}
\Phi -1)]$ has period $4\pi$ and has extrema at $\Phi = 2\pi
p_{1}$, and\, $\Phi = 4\pi p_{2} \pm 2 \mbox{cos}^{-1}
[1-|\mu_{1}/(2\mu_{3})|]$ with $p_{1},p_{2} \in \IZ$; the second
extrema exists only if $|\mu_{1}/(2\mu_{3})|< 1$. From the
mathematical point of view the DSG model belongs to a
 class of theories with partial integrability \cite{weiss}. Depending on the
 values of the parameters $\b_{0},\, \mu_{1},\,\mu_{3}$ the quantum field theory
 version of the DSG model presents a variety of physical effects, such as the decay
 of the false vacuum, a phase transition, confinement of the kinks and the resonance
 phenomenon due to unstable bound states  of excited kink-antikink states (see \cite{mussardo} and
 references therein). The
semi-classical spectrum of neutral particles in the DSG theory is
investigated in \cite{mussardo1}.

Interestingly the functions\footnote{These functions are
 obtained by adding the term $\mbox{exp}[2\g (x - vt)]$ to the relevant tau
 functions for one solitons used above. This procedure adds a new method of solving
 DSG which deserve further study.
 The multi-frequency SG equations can be solved through the Jacobi elliptic function
 expansion method, see e.g.  \cite{liu}.}
 \br\nonumber \tau_{0}&=&1+i\, d\,
\mbox{exp}[\gamma(x-v t)]+ h\, \mbox{exp}[2\gamma(x-vt)],\,\,\,\\
\tau_{1}&=& 1-i\, d\, \mbox{exp}[\gamma(x-v t)]+ h\,
\mbox{exp}[2\gamma(x-vt)],\er satisfy the equation (\ref{dsg}) for
the parameters (\ref{paras}) provided \br \label{exp11} e^{-i
\Phi/2} &=& \tau_{1}/\tau_{0}
\\ \gamma^2&=&\mu_{1}+2\mu_{3},\,\,\, h=
-\frac{\mu_{1}}{4},\,\,\,\, e_{1}=-1\label{h1}\er

The general solution of this type can be written as \br
\label{gen} \Phi := 4\,
\mbox{arctan}\left[\frac{1}{d}\,\,\frac{1+h\,\,\mbox{exp}[2\gamma(x-vt)]}{\mbox{exp}[\gamma(x-vt)]}\right]
\er

\subsubsection{DSG kink ($h < 0, \,\mu_{i}>0$)}

For the choice of parameters $h < 0, \,\mu_{i}>0 $ in (\ref{h1})
the equation (\ref{gen}) provides \br \label{kink} \vp
&=&\frac{4}{\b_{0}} \mbox{arctan}\left[\frac{-2
|h|^{1/2}}{d}\,\,\mbox{sinh}[\gamma_{K}\, (x-vt) +
a_{0}]\right],\,\,\,\,\g_{K}\equiv \pm \sqrt{\mu_{1}+ 2
\mu_{3}},\\
\nonumber && a_{0}=\frac{1}{2} \mbox{ln} |h|. \er

This is the DSG 1-kink solution with mass \br M_{K} =
\frac{16}{\b_{0}^2} \g_{K} \left[1 +\frac{\mu_{1}}{\sqrt{2\mu_{3}
(\mu_{1}+ 2\mu_{3})}}\mbox{ln} (\frac{\sqrt{\mu_{1}+ 2\mu_{3}}+
\sqrt{2\mu_{3}}}{\sqrt{\mu_{1}}})\right]. \er

Notice that in the limit $\mu_{1} \rightarrow 0$ the kink mass
becomes $M_{K} = \frac{16}{\b_{0}^2} \sqrt{2\mu_{3}}$, which is
twice the soliton mass (\ref{massmu3}) of the model (\ref{eqmu10})
for the parameters $\nu_{1}=1/2,\, q=2$. Similarly,  in the limit
$\mu_{3} \rightarrow 0$ the kink  mass becomes
$\frac{8}{(\b_{0}/2)^2} \sqrt{\mu_{1}}$, which is the soliton mass
(\ref{massmu1}) of the model (\ref{eqmu30}) for $\nu_{1}=1/2, \,
q=2$; thus in this case the coupling constant is $\b_{0}/2$. As
discussed  above these solitons get in some sense ``confined''
inside the kink if the parameters satisfy $\mu_{i} \neq 0$. The
1-antikink is plotted in Fig. 4.

\subsubsection{Bounce-like solution  ($h>0$, \,$\mu_{1} < 0$)}

For the parameters $h>0$, \,$\mu_{1} < 0$ one gets from
(\ref{gen}) \br \vp :=\frac{4}{\b_{0}}
\mbox{arctan}\left[\frac{2h^{1/2}}{d}\,\,\mbox{cosh}[\gamma
'(x-vt) + a_{0}^{\prime}]\right],\,\,\,\,\,\g '=  2\mu_{3}-
|\mu_{1}|,\,\,\,\,a_{0}^{\prime}=\frac{1}{2} \mbox{ln} h \er

This is the  bounce-like solution and interpolates between the two
vacuum values $2\pi$  and   $4\pi- 2 \mbox{arcos} (1-
|\mu_{1}/2\mu_{3}|)$ and then it comes back. Since $2\pi$ is a
false vacuum position this solution is not related to any stable
particle in the quantum theory \cite{mussardo}. In Fig. 2 we plot
this profile.

\FIGURE{\centering
\hspace{2.0cm}\scalebox{0.3}{\includegraphics[angle=270]{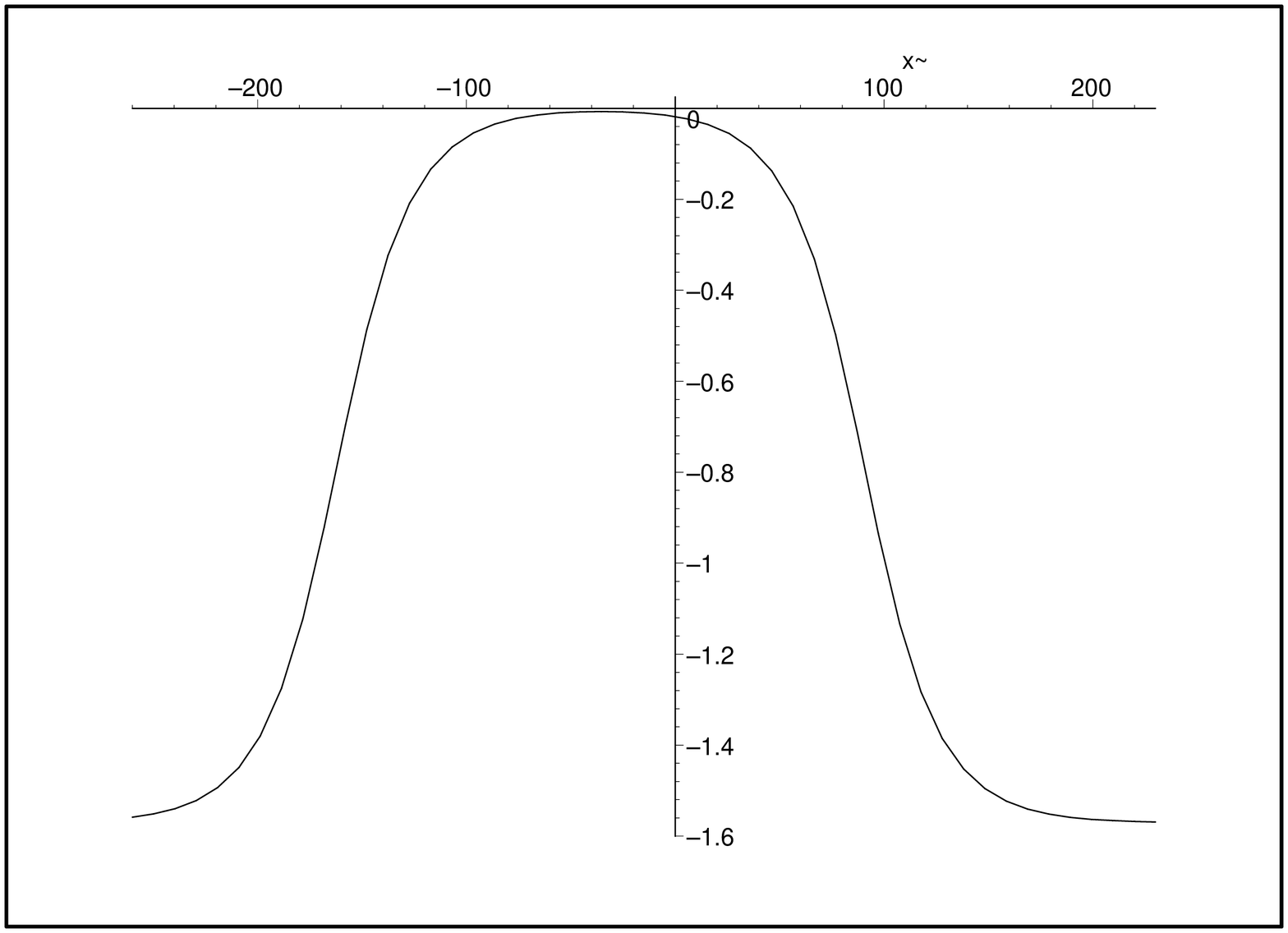}}
\parbox{5in}{\caption{Bounce-like solution ($\frac{\b_{0}}{4} \vp$) plotted for
$\mu_{2}=-0.0000001,\,\mu_{3}=0.001,\,d=-2$.}}}

\newpage

\section{Classical GSG as a reduced affine Toda model coupled to matter}
\label{atm}

In this section we provide the algebraic construction of the
 $sl(3, \IC)$  conformal affine Toda model coupled to matter fields (CATM) and
 closely follows refs. \cite{jhep, bueno, matter}
but the reduction process to arrive at the classical GSG model is
new. The CATM model is a two-dimensional field theory involving
four scalar fields and six Dirac spinors. The interactions among
the fields are as follows: 1) in the scalars equations of motion
there are the coupling of bilinears in the spinors to exponentials
of the scalars. 2) Some of the equations of motion for the spinors
have certain bilinear terms in the spinors themselves. That fact
makes it difficult to find a local Lagrangian for the theory.
Nevertheless, the model presents a lot of symmetries. It is
conformally invariant, possesses local gauge symmetries as well as
vector and axial conserved currents bilinear in the spinors. One
of the most remarkable properties of the model is that it presents
an equivalence between a U(1) vector conserved current, bilinear
in the spinors, and a topological currents depending only on the
first derivative of some scalars. This property allow us to
implement a bag model like confinement mechanism resembling what
one expects to happen in QCD. The model possesses a zero curvature
representation based on the $\hat{sl}3(C)$ affine Kac– Moody
algebra. It constitutes a particular example of the so-called
conformal affine Toda models coupled to matter fields which have
been introduced in \cite{matter}. The corresponding model
associated to $\hat{sl}2(C)$ has been studied in \cite{nucl1}
where it was shown, using bosonization techniques, that the
equivalence between the currents holds true at the quantum level
and so the confinement mechanism does take place in the quantum
theory.

The off-critical affine Toda model coupled to matter (ATM) is
defined
 by gauge fixing the conformal symmetry \cite{jhep, annals}. The previous treatments
of the $sl(3, \IC)$ ATM model used the symplectic and on-shell
decoupling methods to unravel the classical generalized
sine-Gordon  (cGSG) and generalized massive Thirring (GMT) dual
theories describing the strong/weag coupling sectors of the ATM
model \cite{jmp, jhep, annals}. As mentioned above the ATM model
describes some scalars coupled to spinor (Dirac) fields in which
the system of
 equations of motion has a local
gauge symmetry. Conveniently gauge fixing the local symmetry by
setting some spinor bilinears to constants we are able to decouple
the scalar (Toda) fields from the spinors, the final result is a
direct construction of the classical generalized sine-Gordon model
(cGSG) involving only the scalar fields. In the spinor sector we
are left with a system of equations in which the Dirac fields
couple to the cGSG fields. Another instance in which the quantum
version of the generalized sine-Gordon theory arises is in the
process of bosonization of the generalized massive Thirring model
(GMT), which is a multiflavor extension of the usual massive
Thirring model such that, apart from the usual current-current
self-interaction for each flavor, it presents current-current
interactions terms among the various U(1) flavor currents
\cite{epjc}.

The zero curvature condition (see (\ref{zeroc} ) and the Appendix)
gives the following equations of motion for the CATM model
\cite{matter} \br \label{eqnm1} \frac{\partial
^{2}\th_{a}}{4i\,e^{\eta}} &=&m_{1}[e^{\eta
-i\phi_{a}}\widetilde{\psi }_{R}^{l}\psi
_{L}^{l}+e^{i\phi_{a}}\widetilde{\psi }_{L}^{l}\psi
_{R}^{l}]+m_{3}[e^{-i\phi_{3}}\widetilde{\psi } _{R}^{3}\psi
_{L}^{3}+e^{\eta +i\phi_{3}}\widetilde{\psi } _{L}^{3}\psi
_{R}^{3}];\,\,\,\,a=1,2\,\,\,\,\,\,\,\,\,\,\,\,\,\,\,\,\,\,\,
\\
\label{eqnm3} -\frac{\partial ^{2}\widetilde{\nu }}{4}
&=&im_{1}e^{2\eta -\phi_{1}}\widetilde{\psi }_{R}^{1}\psi
_{L}^{1}+im_{2}e^{2\eta -\phi_{2}}\widetilde{\psi }_{R}^{2}\psi
_{L}^{2}+im_{3}e^{\eta -\phi_{3}}\widetilde{\psi } _{R}^{3}\psi
_{L}^{3}+{\bf m}^{2} e^{3\eta },\,\,
\\
\label{eqnm4} -2\partial _{+}\psi _{L}^{1}&=&m_{1} e^{\eta
+i\phi_{1}}\psi _{R}^{1},\,\,\,\,\,\,\,\,\,\,\,\,\,\,\, -2\partial
_{+}\psi _{L}^{2}\,=\,m_{2}e^{\eta +i\phi_{2}}\psi _{R}^{2},
\\
\label{eqnm5} 2\partial _{-}\psi _{R}^{1}&=&m_{1} e^{2\eta
-i\phi_{1}}\psi _{L}^{1}+2i \(\frac{m_{2} m_{3}}{i
m_{1}}\)^{1/2}e^{\eta }(-\psi _{R}^{3} \widetilde{\psi
}_{L}^{2}e^{i\phi _{2}}- \widetilde{\psi }_{R}^{2}\psi
_{L}^{3}e^{-i\phi_{3}}),
\\
\label{eqnm7} 2\partial _{-}\psi _{R}^{2}&=&m_{2} e^{2\eta
-i\phi_{2}}\psi _{L}^{2}+2i\(\frac{m_{1}
m_{3}}{im_{2}}\)^{1/2}e^{\eta }(\psi _{R}^{3} \widetilde{\psi
}_{L}^{1}e^{i\phi _{1}}+ \widetilde{\psi }_{R}^{1}\psi
_{L}^{3}e^{-i\phi _{3}}),
\\
\label{eqnm8} -2\partial _{+}\psi _{L}^{3}&=&m_{3} e^{2\eta +i\phi
_{3}}\psi _{R}^{3}+2i\(\frac{m_{1} m_{2}}{im_{3}}\)^{1/2}e^{\eta
}(-\psi _{L}^{1}\psi _{R}^{2}e^{i\phi_{2}}+\psi _{L}^{2}\psi
_{R}^{1}e^{i\phi _{1}}),
\\
\label{eqnm9} 2\partial _{-}\psi _{R}^{3}&=&m_{3}e^{\eta
-i\phi_{3}}\psi _{L}^{3},\,\,\,\,\,\,\,\,\,\,\,\, 2\partial
_{-}\widetilde{\psi }_{R}^{1}\,=\,m_{1} e^{\eta
+i\phi_{1}}\widetilde{\psi }_{L}^{1},
\\
\label{eqnm10} -2\partial _{+}\widetilde{\psi }_{L}^{1} &=&m_{1}
e^{2\eta -i\phi_{1}}\widetilde{\psi }_{R}^{1}+2i\(\frac{m_{2}
m_{3}}{i m_{1} }\)^{1/2}e^{\eta }(-\psi _{L}^{2}\widetilde{\psi
}_{R}^{3}e^{-i\phi _{3}}-\widetilde{\psi }_{L}^{3}\psi
_{R}^{2}e^{i\phi_{2}}),
\\
\label{eqnm12} -2\partial _{+}\widetilde{\psi }_{L}^{2}
&=&m_{2}e^{2\eta -i\phi_{2}}\widetilde{\psi
}_{R}^{2}+2i\(\frac{m_{1} m_{3}}{i m_{2}} \)^{1/2}e^{\eta }(\psi
_{L}^{1}\widetilde{\psi }_{R}^{3}e^{-i\phi _{3}}+\widetilde{\psi
}_{L}^{3}\psi _{R}^{1}e^{i\phi_{1}}),
\\
\label{eqnm13} 2\partial _{-}\widetilde{\psi }_{R}^{2}&=&m_{2}
e^{\eta+i\phi_{2}}\widetilde{\psi }_{L}^{2},
\,\,\,\,\,\,\,\,\,\,\,\,\,\,\,\, -2\partial _{+}\widetilde{\psi
}_{L}^{3}\,=\,m_{3} e^{\eta -i\phi _{3}}\widetilde{\psi }_{R}^{3},
\\
\label{eqnm15} 2\partial _{-}\widetilde{\psi }_{R}^{3} &=&m_{3}
e^{2\eta +i\phi_{3}}\widetilde{\psi
}_{L}^{3}+2i\(\frac{m_{1}m_{2}}{im_{3}} \)^{1/2}e^{\eta
}(\widetilde{\psi} _{R}^{1}\widetilde{\psi
}_{L}^{2}e^{i\phi_{2}}-\widetilde{\psi }_{R}^{2}\widetilde{\psi
}_{L}^{1}e^{i\phi_{1}}),
\\
\label{eqnm16}
\partial^{2}\eta&=&0,
\er where $\phi_{1}\equiv2 \th_{1}-\th_{2},\,\phi_{2}\equiv
2\th_{2}-\th_{1},\,\phi_{3} \equiv \th_{1}+\th_{2}$. Therefore,
one has \br \phi_{3}=\phi_{1}+\phi_{2}\label{phi123}\er

The $\theta$ fields are considered to be in general complex
fields. In order to define the classical generalized sine-Gordon
model we will consider these fields to be real.

Apart from the {\sl conformal invariance} the above equations exhibit the
$\(U(1)_{L}\)^{2}\otimes \(U(1)_{R}\)^{2}$ {\sl left-right local gauge symmetry} \br
\label{leri1}
\th_{a} &\ra& \th_{a} + \xi_{+}^{a}( x_{+}) + \xi_{-}^{a}( x_{-}),\,\,\,\,a=1,2\\
\widetilde{\nu} &\ra& \widetilde{\nu}\; ; \qquad \eta \ra \eta \\
\psi^{i} &\ra & e^{i( 1+ \gamma_5) \Xi_{+}^{i}( x_{+})
+ i( 1- \gamma_5) \Xi_{-}^{i}( x_{-})}\, \psi^{i},\\
\,\,\,\, \widetilde{\psi}^{i} &\ra& e^{-i( 1+ \gamma_5) (\Xi_{+}^{i})( x_{+})-i ( 1-
\gamma_5) (\Xi_{-}^{i})( x_{-})}\,\widetilde{\psi}^{i},\,\,\, i=1,2,3;\label{leri2}
\\
&&\Xi^{1}_{\pm}\equiv \pm \xi_{\pm}^{2} \mp
2\xi_{\pm}^{1},\,\,\Xi^{2}_{\pm}\equiv \pm \xi_{\pm}^{1}\mp
2\xi_{\pm}^{2},\,\,\Xi_{\pm}^{3}\equiv
\Xi_{\pm}^{1}+\Xi_{\pm}^{2}. \nonumber\er

One can get global symmetries for $\xi_{\pm}^{a}=\mp
\xi_{\mp}^{a}=$ constants. For a model defined by a Lagrangian
these would imply the presence of two vector and two chiral
conserved currents. However, it was found only half of such
currents \cite{bueno}. This is a consequence of the lack of a
Lagrangian description for the $sl(3)^{(1)}$ CATM in terms of the
$B$ and $F^{\pm}$ fields (see Appendix). So, the vector current
\br \label{vec} J^{\mu}= \sum_{j=1}^{3} m_{j}
\bar{\psi}^{j}\gamma^{\mu}\psi^{j}\er and the chiral current \br
\label{chi} J^{5\,\mu} = \sum_{j=1}^{3} m_{j}
\bar{\psi}^{j}\gamma^{\mu}\gamma_{5} \psi^{j}+ 2 \partial_{\mu}
(m_{1}\th_{1}+m_{2} \th_{2})\er are conserved \br
\label{conservation} \pa_{\mu} J^{\mu}=0,\,\,\,\,\,\pa_{\mu}
J^{5\, \mu}=0\er

The conformal symmetry is gauge fixed by setting \cite{annals} \br
\eta = \mbox{const}. \label{eta}\er

The off-critical ATM model obtained in this way exhibits the
vector and topological currents equivalence \cite{matter, annals}
\br \label{equivalence} \sum_{j=1}^{3} m_{j}
\bar{\psi}^{j}\gamma^{\mu}\psi^{j} \equiv \epsilon^{\mu
\nu}\partial_{\nu}
(m_{1}\theta_{1}+m_{2}\theta_{2}),\,\,\,\,\,\,\, m_{3}=m_{1}+
m_{2},\,\,\,\,m_{i}>0. \er

In the next steps we implement the reduction process to get the
cGSG model through a gauge fixing of the ATM theory. The local
symmetries (\ref{leri1})-(\ref{leri2}) can be gauge fixed through
\br \label{gf} i \bar{\psi}^{j}\psi^{j}= i
A_{j}=\mbox{const.};\,\,\,\,\,\,\bar{\psi}^{j}\gamma_{5}\psi^{j}
=0. \er

From the gauge fixing (\ref{gf}) one can write the following
bilinears
 \br
 \label{bilinears}
 \widetilde{\psi}_{R}^{j} \psi_{L}^{j} +
 \widetilde{\psi}_{L}^{j}\psi_{R}^{j}=0,\,\,\,\,\,j=1,2,3;
 \er
so,  the eqs. (\ref{gf})  effectively comprises three gauge fixing
 conditions.

It can be directly verified that the gauge fixing (\ref{gf})
preserves the currents conservation laws (\ref{conservation}),
i.e. from the equations of motion (\ref{eqnm1})-(\ref{eqnm16})
 and the gauge fixing (\ref{gf}) together with (\ref{eta}) it is possible
 to obtain the currents conservation laws (\ref{conservation}).

Taking into account the constraints (\ref{gf}) in the scalar
sector, eqs. (\ref{eqnm1}), we arrive at
 the following system
of equations (set $\eta=0$)\br \label{sys1} \pa^2 \th_{1} &=&
M_{\psi}^{1}\,
\mbox{sin} (2\th_{1}-\th_{2}) + M_{\psi}^{3}\, \mbox{sin} (\th_{1}+\th_{2}),\\
\label{sys2}\,\,\,\pa^2 \th_{2} &=& M^{2}_{\psi}\, \mbox{sin}
(2\th_{2}-\th_{1}) + M^{3}_{\psi}\, \mbox{sin}
(\th_{1}+\th_{2}),\,\,\,\, M^{i}_{\psi} \equiv 4 A_{i}\,
m_{i},\,\,\,\,i=1,2,3.\er

Define the fields $\vp_{1},\,\vp_{2}$ as \br
\label{transf1}\vp_{1} &\equiv& a \th_{1} +
b\th_{2},\,\,\,\,\,\,\,\,
a=\frac{4\nu_{2}-\nu_{1}}{3\b_{0}\nu_{1}\nu_{2}},\,\,\,d=\frac{4\nu_{1}-\nu_{2}}{3\b_{0}\nu_{1}\nu_{2}}
\\\vp_{2}&\equiv& c \th_{1} + d
\th_{2},\,\,\,\,\,\,\,\, b=-c=\frac{2(\n_{1}-\nu_{2})}{3\b_{0}
\nu_{1}\nu_{2}}, \,\,\,\, \nu_{1}, \nu_{2} \in \IR
\label{transf2}\er

Then, the system of equations (\ref{sys1})-(\ref{sys2}) written in
terms of the fields $\vp_{1,\,2}$ becomes \br  \pa^2 \vp_{1} &=& a
M^{1}_{\psi}\, \mbox{sin} [\b_{0}\nu_{1}(2\vp_{1}-\vp_{2})] + b
M^{2}_{\psi}\, \mbox{sin} [\b_{0}\nu_{2}(2\vp_{2}-\vp_{1})] +
\nonumber \\&&(a+b) M^{3}_{\psi}\, \mbox{sin}
\b_{0}[(2\nu_{1}-\nu_{2})\vp_{1}+(2\nu_{2}- \nu_{1})\vp_{2})],
\label{cgsg1} \\
\nonumber \pa^2 \vp_{2} &=& c M^{1}_{\psi}\, \mbox{sin}
[\b_{0}\nu_{1}(2\vp_{1}-\vp_{2})] + d M^{2}_{\psi}\, \mbox{sin}
[\b_{0}\nu_{2}(2\vp_{2}-\vp_{1})] +
\\&& (c+d) M^{3}_{\psi}\, \mbox{sin} \b_{0}[(2\nu_{1}-\nu_{2})\vp_{1}+(2\nu_{2}-
\nu_{1})\vp_{2})]\label{cgsg2} \er

The system of equations above considered for real fields
$\vp_{1,\,2}$ as well as for real parameters $M_{\psi}^{i}, a, b,
c, d, \b_{0}$ defines the {\sl classical generalized sine-Gordon
model} (cGSG).
 Notice that this classical
version of the GSG model derived from the ATM theory is a submodel
of the GSG model (\ref{eq1})-(\ref{eq2}), defined in section
\ref{model}, for the particular parameter values
 $r=\frac{2\nu_{1}-\nu_{2}}{\nu_{3}},\,
 s=\frac{2\nu_{2}-\nu_{1}}{\nu_{3}}$ and the convenient identifications of the
 parameters in the coefficients of the sine functions of the both  models.

The following reduced models can be obtained from the system
(\ref{cgsg1})-(\ref{cgsg2}):

i){\sl SG submodels}\\
i.1) For $\nu_{2}=2\nu_{1}$ \,one has  $M^{1}_{\psi} =
M^{2}_{\psi}$ and the system $\vp_{2}=0$,\,\, $\pa^{2}\vp_{1}=
M^{1}_{\psi} \frac{3\nu_{1}}{\b_{0}}\, \mbox{sin}\,\b_{0} 2\nu_{1}
\vp_{1}$.
\\i.2) For $\nu_{1}=2\nu_{2}$ \,one has  $ M^{1}_{\psi}= M^{2}_{\psi}$
and the system $\vp_{1}=0$,\,\, $\pa^{2}\vp_{2}= M^{2}_{\psi}
\frac{3\nu_{2}}{\b_{0}}\, \mbox{sin}\, \b_{0} 2\nu_{2} \vp_{2}$.
\\ i.3) For $\nu_{2}=\nu_{1}\equiv \nu$\, and\,$\vp_{1}=\vp_{2}\equiv \hat{\vp}_{A},\, (A=1,2) $, one gets the
sub-models

i.3a) $M^{1}_{\psi}=M^{2}_{\psi},\,M^{3}_{\psi}=0$,\,\, $\pa^{2}
\hat{\vp}_{1}= a M^{1}_{\psi} \, \mbox{sin}\, \b_{0} \nu
\hat{\vp}_{1},$

 i.3b) $M^{1}_{\psi}=M^{2}_{\psi}=0$,\,\,\,\,\,\,\,\,\,\,\,\,\,\,\, $\pa^{2} \hat{\vp}_{2}=
a M^{3}_{\psi} \, \mbox{sin}\, \b_{0} \nu \hat{\vp}_{2}.$

ii) {\sl DSG sub-model}\\ For $\nu_{1}=\nu_{2}$ \,and \, $
M^{1}_{\psi}= M^{2}_{\psi}$ one gets the sub-model
$\vp_{1}=\vp_{2}\equiv \vp$,\,\, $\pa^{2}\vp= a M^{1}_{\psi} \,
\mbox{sin}\, \b_{0} \nu_{1} \vp + a M^{3}_{\psi}\, \mbox{sin}\,
2\b_{0} \nu_{1} \vp$.

The sub-models i.1)-i.2) each one contains the ordinary
sine-Gordon model (SG) and they were considered in the subsections
\ref{s11} and \ref{s22}, respectively; the sub-model i.3) supports
two SG models
 with different soliton masses which must correspond to the construction in subsection \ref{s33};
 and the ii) case defines the double
sine-Gordon model (DSG) studied in subsection \ref{dsg:sec}. Other
meaningful reductions are possible arriving at either SG or DSG
model. Notice that the reductions above are particular cases of
the sub-models in subsections \ref{s11}, \ref{s22}, \ref{s33} and
 \ref{dsg:sec}, respectively, for relevant parameter identifications.

In view of the gauge fixing (\ref{gf}) the spinor sector can be
parameterized conveniently as \br \left(
\begin{array}{c}
 \psi_{R}^{j}\\\psi_{L}^{j}
\end{array}\right)= \left(
\begin{array}{c}
\sqrt{A_{j}/2}\,u_{j} \\i \sqrt{A_{j}/2}\,\frac{1}{v_{j}}
\end{array}\right);\,\,\,\,\left(
\begin{array}{c}
 \widetilde{\psi}_{R}^{j}\\\widetilde{\psi}_{L}^{j}
\end{array}\right)= \left(
\begin{array}{c}
\sqrt{A_{j}/2}\, v_{j} \\-i \sqrt{A_{j}/2}\, \frac{1}{u_{j}}
\end{array}\right).\label{paramet}\er

Therefore, in order to find the spinor field solutions one can
solve the eqs. (\ref{eqnm4})-(\ref{eqnm15}) for the fields $u_{j},
v_{j}$ for each solution given for the cGSG fields $\vp_{1,\,2}$
of the system (\ref{cgsg1})-(\ref{cgsg2}).

In the context of the Hirota and dressing transformation methods
 one can construct the soliton solutions of the off-critical ATM model using the fields
parameterizations in terms of the tau functions \cite{bueno} \br
\rme^{-\iu \theta_1} &=& \frac{\hat{\tau}_1}{\hat{\tau}_0}, \qquad
\rme^{-\iu \theta_2} = \frac{\hat{\tau}_2}{\hat{\tau}_0}, \qquad
\rme^{-\({\widetilde \nu} + \frac{1}{12} \sum_{i=1}^3 m_i^2 \, x^+
x^-\)} =
\hat{\tau}_0. \label{phitau}\\
{\psi}_R^1&=& \frac{m_3}{m_1}\frac{\tseventeen}{\tthree}
-\frac{m_2}{m_1} \frac{\tsixteen}{\tone}, \qquad \qquad
{\psi}_L^1=-\frac{\tseven}{\ttwo},
\label{taucampos1} \\
{\psi}_R^2&=& \frac{m_3}{m_2}\frac{\tnineteen}{\ttwo}
-\frac{m_1}{m_2}\frac{\teighteen}{\tone}, \qquad \qquad
{\psi}_L^2=-\frac{\teight}{\tthree},
\\
{\psi}_R^3&=&\frac{\tsix}{\tone}, \qquad \qquad \qquad \qquad
\qquad \quad \; {\psi}_L^3=\frac{m_2}{m_3}
\frac{\tfourteen}{\ttwo}
+\frac{m_1}{m_3}\frac{\tfifteen}{\tthree},
\\
{\tilde \psi}_R^1&=&-\frac{\tfour}{\ttwo}, \qquad \qquad \qquad
\qquad \quad \quad \;\, {\tilde \psi}_L^1=
\frac{m_2}{m_1}\frac{\televen}{\tone}
-\frac{m_3}{m_1}\frac{\tten}{\tthree},
\\
{\tilde \psi}_R^2&=&-\frac{\tfive}{\tthree}, \qquad \qquad \qquad
\qquad \quad \quad \;\, {\tilde \psi}_L^2=
\frac{m_1}{m_2}\frac{\tthirteen}{\tone}
-\frac{m_3}{m_2}\frac{\ttwelve}{\ttwo},
\\
{\tilde\psi}_R^3&=& -\frac{m_2}{m_3}\frac{\ttwenty}{\ttwo}
-\frac{m_1}{m_3}\frac{\ttwentyone}{\tthree}, \qquad \quad \,
{\tilde \psi}_L^3=-\frac{\tnine}{\tone}, \label{taucampos2} \er

These tau functions must satisfy the relationships
 \br \tfive\,\tsix +
\tthree\,\tsixteen - \tone\,\tseventeen &=& 0, \qquad
\tfour\,\tsix - \ttwo\,\teighteen +\tone\,\tnineteen = 0,
\label{algrel1} \\
\tfour\,\tfive - \tthree\,\ttwenty + \ttwo\,\ttwentyone &=& 0,
\qquad
\teight\,\tnine - \tone\,\tten + \tthree\,\televen =0, \\
\tseven\,\tnine + \tone\,\ttwelve - \ttwo\,\tthirteen &=& 0,
\qquad \tseven\,\teight + \tthree\,\tfourteen - \ttwo\,\tfifteen =
0. \label{algrel2} \er

The Hirota method requires the determination of the equations
 of motion satisfied by the {\sl tau} functions. They are determined by substituting
the relations between the fields and tau functions
(\ref{phitau})-(\ref{taucampos2}) into the equations of motion
(\ref{eqnm1})-(\ref{eqnm15}). Except for $\eta$, one has $15$
fields in the model (\ref{eqnm1})-(\ref{eqnm16}) (three scalars
and six two-component spinors). However, in
(\ref{phitau})-(\ref{taucampos2}) we have defined $21$
tau-functions. The algebraic relations
(\ref{algrel1})-(\ref{algrel2}) which arise in the framework of
the dressing method provide the six equations missing. Due to the
local gauge invariance (\ref{leri1})-(\ref{leri2}) from the 15
fields of the model mentioned above only 9 of them describe the
physical degrees of freedom which will be related to the physical
soliton spectrum of the theory as discussed below.

The dressing transformation method does not excite the field
$\eta$ if one starts from a solution where it vanishes. Moreover,
it has been shown that the soliton type solutions are in the orbit
of the vacuum $\eta=0$ \cite{matter}.

\subsection{Physical solitons and kinks of the ATM model}

The main feature of the one `solitons' constructed in \cite{bueno}
is that for each positive root of $sl(3)$ there corresponds one
soliton species associated to the fields
$\phi_{1},\,\phi_{2},\,\phi_{3}$, respectively. The relevant
solutions for the spinor fields together with the 1-`solitons'
satisfy the relationship (\ref{equivalence}). The class of
$2$-`soliton' solutions of $sl(3)$ ATM obtained in \cite{bueno}
behave as follows:\, i) they are given by 6 species associated to
the pair $(\a_{i},\a_{j}),\, i\le j;\,\, i,j=1,2,3$; where the
$\a$'s are the positive roots of $sl(3)$ Lie algebra. Each species
$(\a_{i},\a_{i})$ solves the $sl(2)$ ATM submodel\footnote{$sl(2)$
ATM gauge unfixed $2-$'solitons' satisfy an analogous eq. to
\ref{equivalence}. Moreover, for $\vp$ real and
$\widetilde{\psi}=\pm (\psi)^{*}$ one has, soliton-soliton $SS$,
$SS$ bounds and no $S\bar{S}$ ($S=$soliton,\,
$\bar{S}$=anti-soliton) bounds \cite{nucl1} associated to the
field $\vp$.}. ii) they satisfy the $U(1)$ vector and topological
currents equivalence \ref{equivalence}. However, the possible kink
type solutions associated in a {\sl non-local} way to the spinor
bilinears and the relevant gauge fixing of the local symmetry
(\ref{leri1})-(\ref{leri2}) have not been discussed in the
literature. In order to consider the physical spectrum of solitons
and study their properties, such as masses and scattering time
delays, it is mandatory to take into account these questions which
are related to the counting of the true physical degrees of
freedom of the theory. Therefore, one must consider the possible
soliton type solutions  associated to each spinor bilinear. The
relation between  this type of `solitons', say $\hat{\phi}_{j}$,
and their relevant fermion bilinears must be non-local as
suggested by the equivalence equation (\ref{equivalence}). So, we
may have soliton solutions of type \br \hat{\phi}_{j} = \int^{x}\,
dx^{\prime} \, \bar{\psi}^{j}\g^{0}\psi^{j} ,\,\,\,\,\,\,j=1,2,3
\label{nonlocal}\er

At this stage one is able to enumerate the physical 1-soliton
(1-antisoliton) spectrum associated to the gauge fixed ATM model.
In fact, we have three 'kinks' and their corresponding
'anti-kinks' associated to the fields $\phi_{i}$ (i=1,2,3), and
three kink and antikink pairs of type $\hat{\phi}_{j},\,j=1,2,3$.
Thus, we have six kink and their relevant antikink solutions, but
in order to record the physical soliton and anti-soliton
excitations one must take into account the four constraints
(\ref{phi123}) and (\ref{bilinears}). Therefore, we expect to find
four pairs of soliton and anti-soliton physical excitations in the
spectrum. This feature is nicely reproduced in the cGSG sector of
the ATM model; in fact, in the last section we were able to write
four usual sine-Gordon models as possible reductions of the cGSG
model. Namely, 1-soliton (1-antisoliton) associated to the fields
$\vp_{1},\, \vp_{2}$, respectively (subsections \ref{s11} and
\ref{s22}) and 1-solitons (1-antisolitons) associated to the
fields $\hat{\vp}_{A},\,A=1,2$, respectively (subsection
\ref{s33}), thus, providing four pairs of 1-
soliton/1-antisoliton. In the 2-kink (2-antikink) sector a similar
argument will provide us ten physical 2-solitons and their
relevant 2-antisoliton excitations, i.e. six pairs of 2-kink and
2-antikink solutions of type $\phi$
 and $\hat{\phi}$, respectively, which give twenty four excitations, and
 taking into account  the
 constraints (\ref{phi123}) and (\ref{bilinears}) we are left with
 ten pairs of 2-solitons and 2-antisolitons. In fact, these
 ten 2-solitons correspond to the pairs we can form with the four
 species of 1-solitons found in the last section, in all possible
 ways. The same argument holds for the corresponding ten
 2-antisolitons.

 In this way the system (\ref{cgsg1})-(\ref{cgsg2}) gives rise to a
 richer (anti)soliton spectrum and dynamics than the $\theta_{a}$ field
'soliton' type solutions  of the gauge unfixed model
(\ref{eqnm1})-(\ref{eqnm15}) found in \cite{bueno}. Regarding this
issue let us notice that in the procedure followed in ref.
\cite{bueno} the local symmetry (\ref{leri1})-(\ref{leri2}) and
the relevant gauge fixing has not been considered explicitly,
therefore their 'solitons' do not correspond to the GSG solitons
obtained above.

Notice that the tau functions in section \ref{model} possess the
function $\g (x-v t)$ in their exponents, whereas the
corresponding ones in the ATM theory  have two times this function
\cite{bueno, nucl1}. This fact is reflected in the GSG soliton
solutions which are two times the relevant solutions of the ATM
model.  It has been observed already in the $sl(2)$ case that the
 $\theta$ `soliton' of the gauge unfixed $sl(2)$ ATM model (see eq. (2.22) of
\cite{nucl1}) is half the soliton of the usual SG model.

\section{Topological charges, baryons as solitons and confinement}
 \label{topological}

In this section we will examine the vacuum configuration of the
cGSG model and the equivalence between the $U(1)$ spinor current
and  the topological current (\ref{equivalence}) in the gauge
fixed model and verify that the charge associated to the $U(1)$
current gets confined inside the solitons and kinks of the GSG
model obtained in section \ref{model}.

It is well known that in 1 + 1 dimensions the topological current
is defined as $J_{\mbox{top}}^{\mu} \sim
\epsilon^{\mu\nu}\pa_{\nu}\Phi$, where $\Phi$ is some scalar
field. Therefore, the topological charge is $ Q_{\mbox{top}} =
\int J_{\mbox{top}}^{0} dx \sim  \Phi(+ \infty) - \Phi(-\infty) $.
In order to introduce a topological current we follow the
construction adopted in Abelian  affine Toda models, so we define
the field \br \theta = \sum_{a=1}^{2} \frac{2 \a_{a}}{\a^2_{a}}
\theta_{a} \er where $\a_{a},\,a=1,2$, are the simple roots of
$sl(3, \IC)$. We then have that $\theta_{a} = (\theta | \l_{a})$,
where $\l_{a}$ are the fundamental weights of $sl(3, \IC)$ defined
by the relation \cite{humphreys} \br 2 \frac{( \a_{a} |
\l_{b})}{(\a_{a}|\a_{a})}= \d_{ab}. \er

The exponentials in the fields $\phi_{j}$ in the equations
(\ref{eqnm1})-(\ref{eqnm15}) written as the combinations $(\theta|
\a_{j}), \, j=1,2,3$, where the $\a_{j}'s$ are the positive roots
of  $sl(3, \IC)$, are invariant under the transformation \br
\theta \rightarrow \theta + 2\pi
\mu\,\,\,\,\,\,\,\,&\mbox{or}&\,\,\,\,\,\,\,\,\phi_{j}\rightarrow
\phi_{j} + 2\pi
(\mu|\a_{j}),\label{transdiscret}\\
\mu &\equiv& \sum_{n_{a}\in \IZ} n_{a} \frac{2
\vec{\l}_{a}}{(\a_{a}|\a_{a})},\label{weight}\er where $\mu$ is a
weight vector of $sl(3, \IC)$, these vectors satisfy $(\mu
|\a_{j}) \in \IZ $ and form an infinite discrete lattice called
the weight lattice \cite{humphreys}. However, this weight lattice
does not constitute the vacuum configurations of the ATM model ,
since in the model described by (\ref{eqnm1})-(\ref{eqnm16}) for
any constants $\theta_{a}^{(0)}$ and $\eta^{(0)}$ \br
\label{vacuum} \psi_{j} = \widetilde{\psi}_{j}
=0,\,\,\theta_{a}=\theta_{a}^{(0)},\,\,\eta=\eta^{(0)},\,\,\,\widetilde{\nu}
= - {\bf{m}}^{2}e^{\eta^{(0)}} x^{+} x^{-}\er is a vacuum
configuration.

We will see that the topological charges of the physical
one-soliton solutions of (\ref{eqnm1})-(\ref{eqnm16}) which are
associated to the new fields $\vp_{a},\,a=1,2,$ of the cGSG model
(\ref{cgsg1})-(\ref{cgsg2}) lie on a modified lattice which is
related to the weight lattice by re-scaling the weight vectors. In
fact, the eqs. of motion (\ref{cgsg1})-(\ref{cgsg2}) for the field
defined by $\vp \equiv \sum_{a=1}^{2} \frac{2 \a_{a}}{\a^2_{a}}
\vp_{a},\,$ such that $\vp_{a} = (\vp | \l_{a})$, are invariant
under the transformation \br \vp \rightarrow \vp +
\frac{2\pi}{\b_{0}}\sum_{a=1}^{2}
 \frac{q_{a}}{\nu_{a}} \frac{2
\l_{a}}{(\a_{a}|\a_{a})},\,\,\,q_{a} \in \IZ.\er

So, the vacuum configuration is formed by an infinite discrete
lattice related to the usual weight lattice by the relevant
re-scaling of the fundamental weights $\l_{a}\rightarrow
\frac{1}{\nu_{a}} \l_{a}$. The vacuum lattice can be given by the
points in the plane $\vp_{1}$\, x\, $\vp_{2}$ \br (\vp_{1}
\,,\,\vp_{2}) = \frac{2\pi}{3\b_{0}}(\frac{2 q_{1}}{\nu_{1}} +
\frac{ q_{2}}{\nu_{2}}\,,\,\frac{ q_{1}}{\nu_{1}}+\frac{2
q_{2}}{\nu_{2}}),\,\,\,\,\,q_{a} \in \IZ.\er

In fact, this lattice is related to the one in eq.
(\ref{lattice1}) through appropriate parameter identifications. We
shall define the topological current and charge, respectively, as
\br
 J_{\mbox{top}}^{\mu} = \frac{\b_{0}}{2\pi} \epsilon^{\mu\nu} \pa_{\nu}
 \vp,\,\,\,\,\,\,\,\,
 Q_{\mbox{top}} = \int dx J_{\mbox{top}}^{0} = \frac{\b_{0}}{2\pi}
 [\vp (+ \infty) -\vp(\infty)]. \label{topcurrcgsg}\er

Taking into account the cGSG fields (\ref{cgsg1})-(\ref{cgsg2})
and the spinor parameterizations (\ref{paramet}) the currents
equivalence (\ref{equivalence}) of the  ATM model takes the form
 \br
\label{equivalence1} \sum_{j=1}^{3} m_{j}
\bar{\psi}^{j}\gamma^{\mu}\psi^{j} \equiv \epsilon^{\mu
\nu}\partial_{\nu} ({\zeta}^{1}_{\psi}\,
\vp_{1}+\zeta^{2}_{\psi}\, \vp_{2}), \er where $\zeta^{1}_{\psi}
\equiv \b_{0}^{2} \nu_{1}\nu_{2} (m^{1}_{\psi} d + m^{2}_{\psi}
b),\,\,\, \zeta^{2}_{\psi} \equiv \b_{0}^{2}\nu_{1}\nu_{2} (m_{2}
a -  m_{1} b)$, $a, b, d$ are defined in
(\ref{transf1})-(\ref{transf2}), and the spinors are understood to
be written in terms of the fields $u_{j}$\, and \,$v_{j}$ of
(\ref{paramet}).

Notice that the topological current in (\ref{equivalence1}) is the
projection of (\ref{topcurrcgsg}) onto the vector
$\frac{2\pi}{\b_{0}} \( \zeta_{\psi}^{1} \l_{1} + \zeta_{\psi}^{2}
\l_{2}\)$.

As mentioned in section \ref{atm} the gauge fixing (\ref{gf})
preserves the currents conservation laws (\ref{conservation}).
Moreover, the cGSG model was defined for the off critical ATM
model obtained after setting $\eta=\mbox{const}.=0$. So, for the
gauge fixed model it is expected to hold the currents equivalence
relation (\ref{equivalence}) written for the spinor
parameterizations  $u_{j}, v_{j}$ and the fields $\vp_{1,2}$ as is
presented in eq. (\ref{equivalence1}). Therefore, in order to
verify the $U(1)$ current confinement it is not necessary to find
the explicit solutions for the spinor fields. In fact, one has
that the current components are given by relevant partial
derivatives of the linear combinations of the field solutions,
$\vp_{1, 2}$, i.e. \, $J^{0}=\sum_{j=1}^{3} m_{j}
\bar{\psi}^{j}\gamma^{0}\psi^{j} \, = \, \partial_{x}
({\zeta}^{1}_{\psi}\, \vp_{1}+\zeta^{2}_{\psi}\,
\vp_{2})$\,and\,$J^{1}=\sum_{j=1}^{3} m_{j}
\bar{\psi}^{j}\gamma^{1}\psi^{j} \, = \, -\partial_{t}
({\zeta}^{1}_{\psi}\, \vp_{1}+\zeta^{2}_{\psi}\, \vp_{2})$. In
particular the current components $J^{0}, J^{1}$ and their
associated scalar field solutions are depicted in Figs. 3 and 4,
respectively, for SG antisoliton and DSG antikink solutions.

It is clear that the charge density related to this $U(1)$ current
can only take significant values on those regions where the
$x-$derivative of the fields $\vp_{1,2}$ are non-vanishing. That
is one expects to happen with the bag model like confinement
mechanism in quantum chromodynamics (QCD). As we have seen the
soliton and kink solutions of the GSG theory are localized in
space, in the sense that the scalar fields interpolate between the
relevant vacua in a limited region of space with a size determined
by the soliton masses. The spinor $U(1)$ current gets the
contributions from all the three spinor flavors. Moreover, from
the equations of motion (\ref{eqnm4})-(\ref{eqnm15})
 one can obtain nontrivial spinor solutions different from vacuum
(\ref{vacuum}) for each set of scalar field solutions $(\vp_{1},
\vp_{2})$. For example, the solution $\vp_{1}=$soliton,
$\vp_{2}=0$ in section \ref{s11} implies $\phi_{1}=\vp_{1},\,
\phi_{2}=-\vp_{1},\,\phi_{3}=0$ which substituting into the spinor
equations of motion (\ref{eqnm4})-(\ref{eqnm15}) will give
nontrivial spinor field solutions. Therefore, the ATM model of
section \ref{atm} can be considered as a multiflavor
generalization of the two-dimensional hadron model proposed in
\cite{chang, Uchiyama}. In the last reference a scalar field is
coupled to a spinor such that the
 DSG kink arises as a model for hadron and the quark field
 is confined inside the bag realizing some properties of
 the MIT bag model.

In connection to our developments above let us notice that
two-dimensional QCD$_{2}$ has been used as a laboratory for
studying the full four-dimensional theory providing an explicit
realization of baryons as solitons. It has been conjectured that
the low-energy action of QCD$_{2}$ ($e
>> m_{q}$, $m_{q}$ quark mass and $e$ gauge coupling) might be related to
massive two dimensional integrable models, thus leading to the
exact solution of the strong coupled QCD$_{2}$ \cite{frishman}.
The baryons in QCD$_{2}$ may be described as solitons in a
bosonized formulation. In the strong-coupling limit the static
classical soliton which describes a baryon in QCD$_{2}$ turns out
to be the sine-Gordon soliton. In particular, it has been shown
that the $sl(2)$ ATM model describes the low-energy spectrum of
QCD$_{2}$ (1 flavor and $N_{c}$ colors) and the exact computation
of the string tension was performed \cite{prd}. A key role has
been played by the equivalence between the Noether and topological
currents at the quantum level. Moreover, one notice that the
SU$(n)$ ATM theory \cite{jmp, jhep} is a $2D$ analogue of the
chiral quark soliton model proposed to describe solitons in
QCD$_{4}$ \cite{diakonov}, provided that the pseudo-scalars lie in
the Abelian subalgebra and certain kinetic terms are supplied for
them.
\newpage
\FIGURE{\centering
\hspace{1.0cm}\scalebox{0.4}{\includegraphics[angle=0]{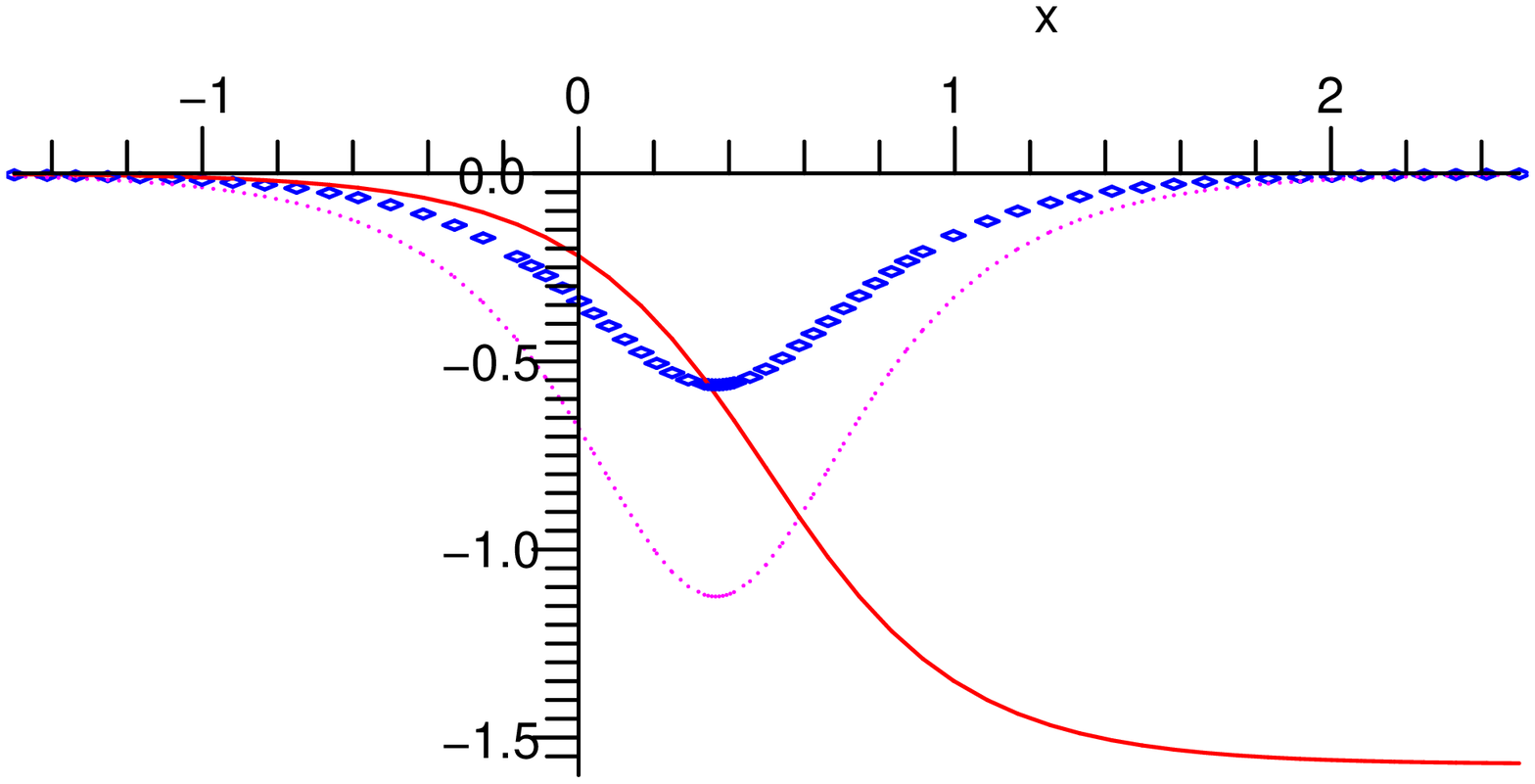}}
\parbox{5in}{\caption{1-antisoliton and
confined current $J^{\mu}$. The solid curve is the 1-antisoliton
($\frac{\b_{0}}{4} \vp$), the dashdotted curve is $J^{0}$ and the
curve with losangles is $J^{1}$. For $t=1,\,\mu_{1}=\mu_{2}=1,
d=1.5, v=0.05, \b_{0}=0.5,\,
m_{\psi}^{1}=m_{\psi}^{2}=1$,\,$\nu_{1}=1,\,\d_{1}=1,\,\d_{2}=2$.}}}

\vskip 1.0cm

\FIGURE{\centering
\hspace{2.0cm}\scalebox{0.4}{\includegraphics{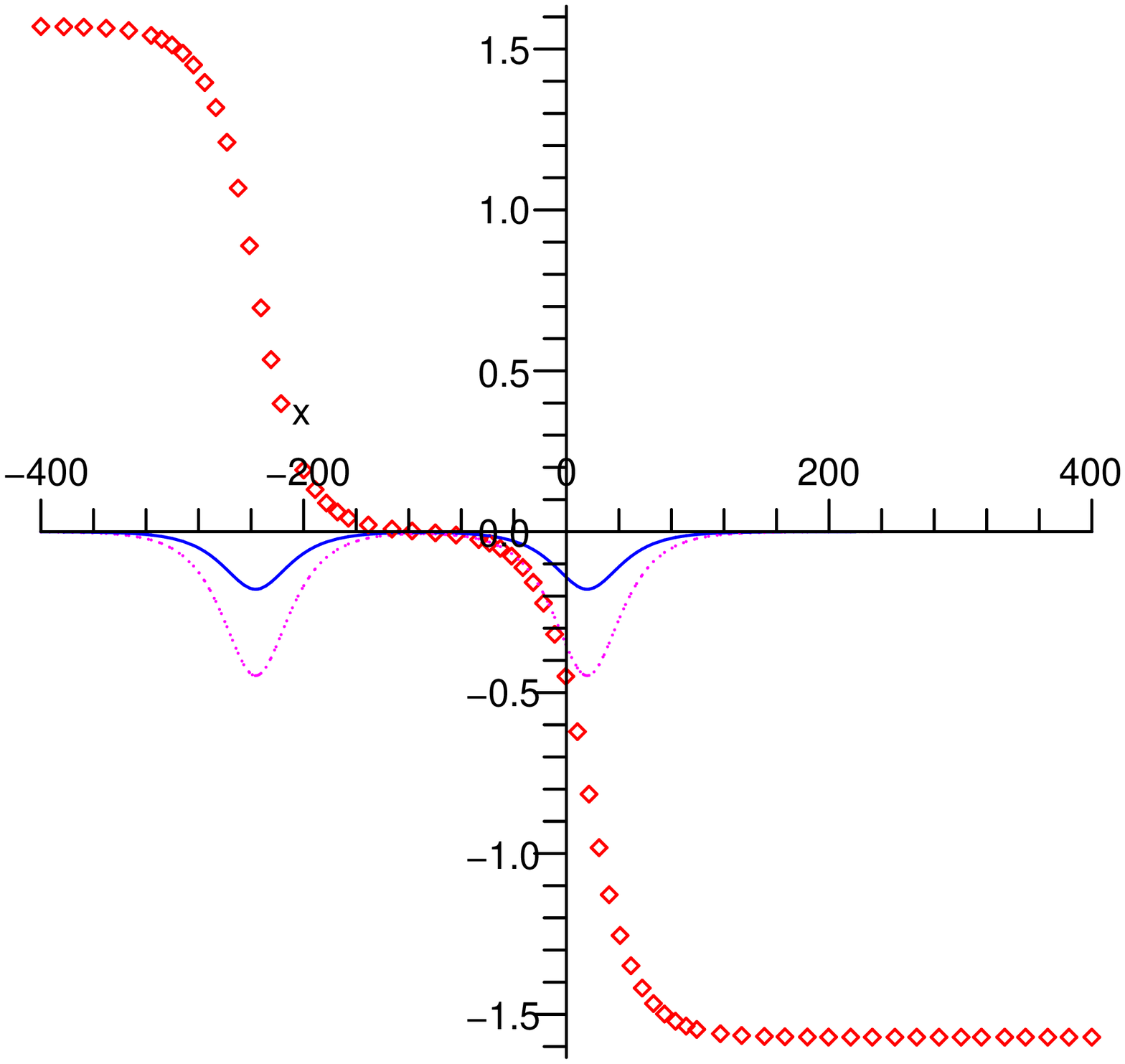}}
\parbox{5in}{\caption{DSG kink solution and confined current $J^{\mu}$. The curve with
losangles is the antikink ($\frac{\b_{0}}{4} \vp$), the dashdotted
line is $J^{0}$, the solid curve is $J^{1}$. For $t=1,\,\b_{0}=
10^8,\, m^{1,
2}_{\psi}=\mu_{1}=-0.0000001,\,\,\mu_{3}:=0.001,\,\,d=2,\,\d_{1}=\d_{2}=1,\,\nu_{1}=1/2
$.}}}

\newpage
\section{Discussion}

The generalized sine-Gordon model GSG (\ref{eq1})-(\ref{eq2})
 provides a variety of solitons, kinks and bounce type solutions. The appearance
of the non-integrable double sine-Gordon model as a sub-model of
the GSG model suggests that this model is a non-integrable theory
for the arbitrary set of values of the parameter space. However, a
subset of values in parameter space determine some reduced
sub-models which are integrable, e.g. the sine-Gordon submodels of
subsections \ref{s11}, \ref{s22} and \ref{s33}.

In connection to the ATM spinors it was suggested that they are
confined inside the GSG solitons and kinks since the gauge fixing
procedure does not alter the $U(1)$ and topological currents
equivalence (\ref{equivalence}). Then, in order to observe the bag
model confinement mechanism it is not necessary to solve for the
spinor fields since it naturally arises from the currents
equivalence relation. In this way our model presents a bag model
like confinement mechanism as is expected in QCD.

The (generalized) massive Thirring model (GMT) is bosonized to the
GSG model \cite{epjc}, therefore, in view of the
 solitons and kinks found above as solutions of the GSG model we expect
 that the spectrum of the GMT model will contain $4$ solitons and their
 relevant anti-solitons, as well
as the kink and antikink excitations. The GMT Lagrangian describes
three flavor massive spinors with current-current interactions
among themselves. So, the total number of solitons which appear in
the bosonized sector suggests that the additional soliton
(fermion) is formed due to the interactions between the currents
in the GMT sector. However, in subsection \ref{s33} the soliton
masses $M_{3}$ and $M_{4}$ become the same for the case
$\mu_{1}=\mu_{2}$, consequently, for this case we have just three
solitons in the GSG spectrum, i.e., the ones with masses $M_{1}$,
$M_{2}$ (subsections \ref{s11}-\ref{s22} ) and $M_{3}=M_{4}$
(subsection \ref{s33}), which will correspond in this case to each
fermion flavor of the GMT model. Moreover, the $sl(3,\IC)$ GSG
model potential (\ref{potential}) has the same structure as the
effective Lagrangian of the massive Schwinger model with $N_{f}=3$
fermions,
 for a convenient value of the vacuum angle $\theta$. The
multiflavor Schwinger model resembles with four-dimensional QCD in
many respects (see e.g. \cite{hosotani} and references therein).

The $sl(n,\IC)$ ATM models may be relevant in the construction of
the low-energy effective theories of multiflavor QCD$_{2}$ with
the dynamical fermions in the fundamental and adjoint
representations. Notice that in the  ATM models the Noether and
topological currents and the generalized sine-Gordon/massive
Thirring models equivalences take place at the classical
\cite{jhep, annals} and quantum mechanical level \cite{epjc,
nucl1}.

Moreover, the interest in baryons with 'exotic' quantum numbers
has recently been stimulated by various reports of baryons
composed by four quarks and an antiquark. The existence of these
baryons cannot yet be regarded as confirmed, however, reports of
their existence have stimulated new investigations about baryon
structure (see e.g. \cite{kabana} and references therein).
Recently, there is new strong evidence of an extremely narrow
$\Theta^{+}$ resonance from DIANA collaboration and a very
significant new evidence from LEPS. According to Diakonov, ``the
null results from the new round of CLAS experiments are compatible
with what one should expect based on the estimates of production
cross sections" \cite{diakonov1}.

Finally, the spectrum of exotic baryons in QCD$_{2}$, with
$SU(N_{f})$ flavor symmetry, has been discussed providing strong
support to the chiral-soliton picture for the structure of normal
and exotic baryons in four dimensions \cite{ellis}. The new
puzzles in non-perturbative QCD are related to systems with
unequal quark masses, so the QCD$_{2}$ calculation must take into
account the $SU(N_{f})$-breaking mass effects, i.e. for $N_{f}=3$
it must be $m_{s} \neq m_{u, d}$. So, in view of our results
above, the properties of the GSG  and the ATM theories may find
some applications in the study of mass splitting of baryons in
QCD$_{2}$ and the understanding of the internal structure of
baryons; a work in this direction is under current research
\cite{hb}.

\vskip 1.0cm

 {\sl\ Acknowledgements}

HB thanks the Mathematics and Physics Departments-UFMT (Cuiab\'a) and IMPA (Rio
de Janeiro) for hospitality and CNPq-FAPEMAT for support. HLC
thanks IMPA for invitation to the ICMP-2006 and FAPESP for
support.

\appendix

\section{The zero-curvature formulation of the CATM model}
\label{atmapp}

We summarize the  zero-curvature formulation of the $sl(3)$ CATM
model \cite{jmp, jhep, bueno}. Consider the zero curvature
condition \br \label{zeroc}
\partial_{+}A_{-}-\partial _{-}A_{+}+[A_{+},A_{-}]=0.
\er

The potentials take the form \br \label{aa1} A_{+}=-B
F^{+}B^{-1},\quad A_{-}=-\partial _{-}BB^{-1}+F^{-},\qquad \er
with \br \label{aa2} F^{+} \,=\,F_{1}^{+}+F_{2}^{+},\,\,\,\,\,\,
F^{-} \,=\, F_{1}^{-}+F_{2}^{-}, \er where $B$ and $F_{i}^{\pm }$
contain the fields of the model \br \label{F1}
F_{1}^{+}&=&\sqrt{im_{1}}\psi _{R}^{1}E_{\alpha
_{1}}^{0}+\sqrt{im_{2}}\psi _{R}^{2}E_{\alpha
_{2}}^{0}+\sqrt{im_{3}}\widetilde{\psi }_{R}^{3}E_{-\alpha
_{3}}^{1},
\\
\label{F2} F_{2}^{+}&=&\sqrt{im_{3}}\psi _{R}^{3}E_{\alpha
_{3}}^{0}+\sqrt{im_{1}} \widetilde{\psi
}_{R}^{1}E_{-\alpha_{1}}^{1}+\sqrt{im_{2}}\widetilde{\psi }
_{R}^{2}E_{-\alpha _{2}}^{1},
\\
\label{3} F_{1}^{-}&=&\sqrt{im_{3}}\psi _{L}^{3}E_{\alpha
_{3}}^{-1}+\sqrt{im_{1}} \widetilde{\psi }_{L}^{1}E_{-\alpha
_{1}}^{0}+\sqrt{im_{2}}\widetilde{\psi } _{L}^{2}E_{-\alpha
_{2}}^{0},
\\
\label{F4} F_{2}^{-}&=&\sqrt{im_{1}}\psi _{L}^{1}E_{\alpha
_{1}}^{-1}+\sqrt{im_{2}}\psi _{L}^{2}E_{\alpha
_{2}}^{-1}+\sqrt{im_{3}}\widetilde{\psi }
_{L}^{3}E_{-\alpha _{3}}^{0},\\
B&=&e^{i\th_{1} H^{0}_{1}+i\th_{2} H^{0}_{2} }\,e^{\widetilde{\nu }C}\,e^{\eta
  Q_{ppal}}\equiv b\, e^{\widetilde{\nu }C}\,e^{\eta
  Q_{ppal}}. \label{equn1}
\er

$E_{\alpha _{i}}^{n},H^{n}_{1},H^{n}_{2}$ and  $C$ ($i=1,2,3; \, n=0,\pm 1$) are some
generators of $sl(3)^{(1)}$; $Q_{ppal}$ being the principal gradation operator. The
commutation relations for an affine Lie algebra in the Chevalley basis are \br
&&\left[ \emph{H}_a^m,\emph{H}_b^n\right] =mC\frac{2}{\alpha_{a}^2}K_{a b}\delta _{m+n,0}  \label{a7}\\
&&\left[ \emph{H}_a^m,E_{\pm \alpha}^n\right] = \pm K_{\alpha a}E_{\pm \alpha}^{m+n}
\label{a8}\\
&&\left[ E_\alpha ^m,E_{-\alpha }^n\right] =\sum_{a=1}^rl_a^\alpha
\emph{H}_a^{m+n}+\frac 2{\alpha ^2}mC\delta _{m+n,0}  \label{a9}
\\
&&\left[ E_\alpha ^m,E_\beta ^n\right] = \varepsilon (\alpha
,\beta )E_{\alpha +\beta }^{m+n};\qquad \mbox{if }\alpha +\beta
\mbox{ is a root \qquad }  \label{a10}
\\
&&\left[ D,E_\alpha ^n\right] =nE_\alpha ^n,\qquad \left[ D,\emph{H}%
_a^n\right] =n\emph{H}_a^n.  \label{a12} \er where $K_{\alpha
a}=2\a.\a_{a}/\a_{a}^2=n_{b}^{\a}K_{ba}$, with $n_{a}^{\a}$ and
$l_a^\alpha$ being the integers in the expansions
$\a=n_{a}^{\a}\a_{a}$ and $\a/\a^2=l_a^\alpha\a_{a}/\a_{a}^2$, and
$\varepsilon (\alpha ,\beta )$ the relevant structure constants.

Take $K_{11}=K_{22}=2$ and $K_{12}=K_{21}=-1$ as the Cartan matrix elements of the
simple Lie algebra $sl(3)$. Denoting by $\a_{1}$ and $\a_{2}$ the simple roots and
the highest one by $\psi (=\a_{1}+\a_{2})$, one has $l_{a}^{\psi}=1(a=1,2)$, and
$K_{\psi 1}=K_{\psi 2}=1$. Take $\varepsilon (\alpha ,\beta )=-\varepsilon (-\alpha
,-\beta ),\,\, \varepsilon_{1,2}\equiv \varepsilon (\alpha_{1} ,\a_{2})=1,\,\,
\varepsilon_{-1,3}\equiv \varepsilon(-\alpha_{1} ,\psi )=1\,\, \mbox{and}\,
\,\,\varepsilon_{-2,3}\equiv \varepsilon (-\alpha_{2} ,\psi)=-1$.

One has $Q_{ppal} \equiv \sum_{a=1}^{2}  {\bf s}_{a}\l^{v}_{a}.H +
3 D$, where $\l^{v}_{a}$ are the fundamental co-weights of
$sl(3)$, and the principal gradation vector is ${\bf s}=(1,1,1)$
\cite{kac}.

\end{document}